%% file: main.tex
\documentclass[10pt,twocolumn,letterpaper]{article}

\usepackage[pagenumbers]{cvpr} 

\input{preamble}

\definecolor{tableblue}{RGB}{225,245,245}
\usepackage[pagebackref,breaklinks,colorlinks,citecolor=TealBlue,urlcolor=magenta]{hyperref}
\usepackage[capitalize]{cleveref}

\usepackage[accsupp]{axessibility} 

\def\paperID{4310} 
\def\confName{CVPR}
\def\confYear{2024}

\begin{document}
\title{CAMixerSR: Only Details Need More ``Attention"}

\author{
	Yan Wang\textsuperscript{\rm 1,2}\footnotemark[3] \authorskip
        Yi Liu\textsuperscript{\rm 1}\footnotemark[1]  \authorskip
	Shijie Zhao\textsuperscript{\rm 1}\footnotemark[1]\hspace{0.3mm}~\footnotemark[2]  \authorskip
	Junlin Li\textsuperscript{\rm 1} \authorskip 
	Li Zhang\textsuperscript{\rm 1}
        \\
	\textsuperscript{\rm 1}Bytedance Inc. \authorskip
	\textsuperscript{\rm 2}Nankai-Baidu Joint Lab, Nankai University 
   \\
 {\tt\small wyrmy@foxmail.com\, \{zhaoshijie.0526,\,liuyi.chester,\,lijunlin.li,\,lizhang.idm\}@bytedance.com}
}

\twocolumn[{%
\renewcommand\twocolumn[1][]{#1}%
\maketitle
\begin{center} 
\centering 
\vspace{-4mm}
\begin{tabular}{c:c:c}
\includegraphics[height=0.335\textwidth]{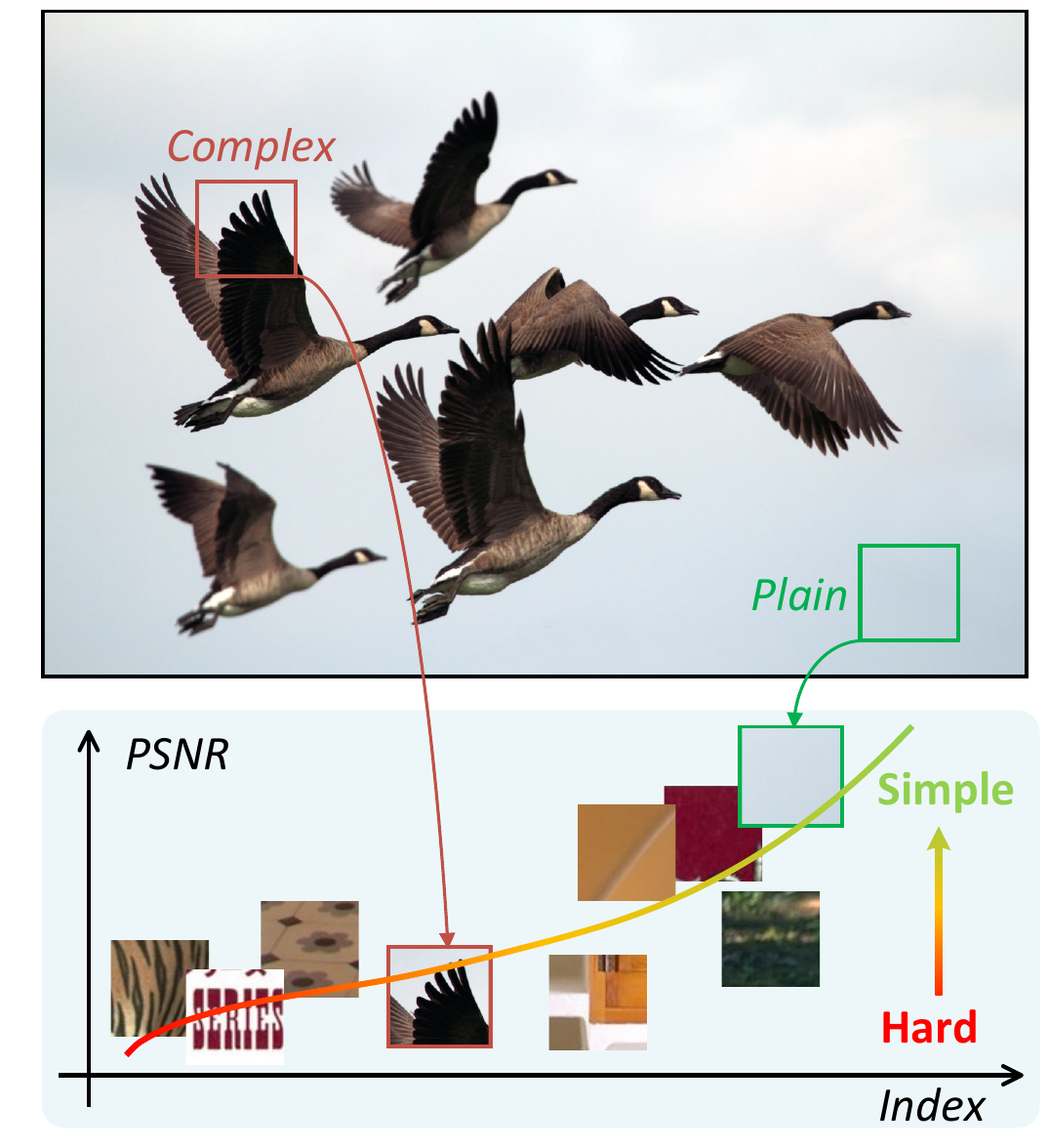}
& \includegraphics[height=0.335\textwidth]{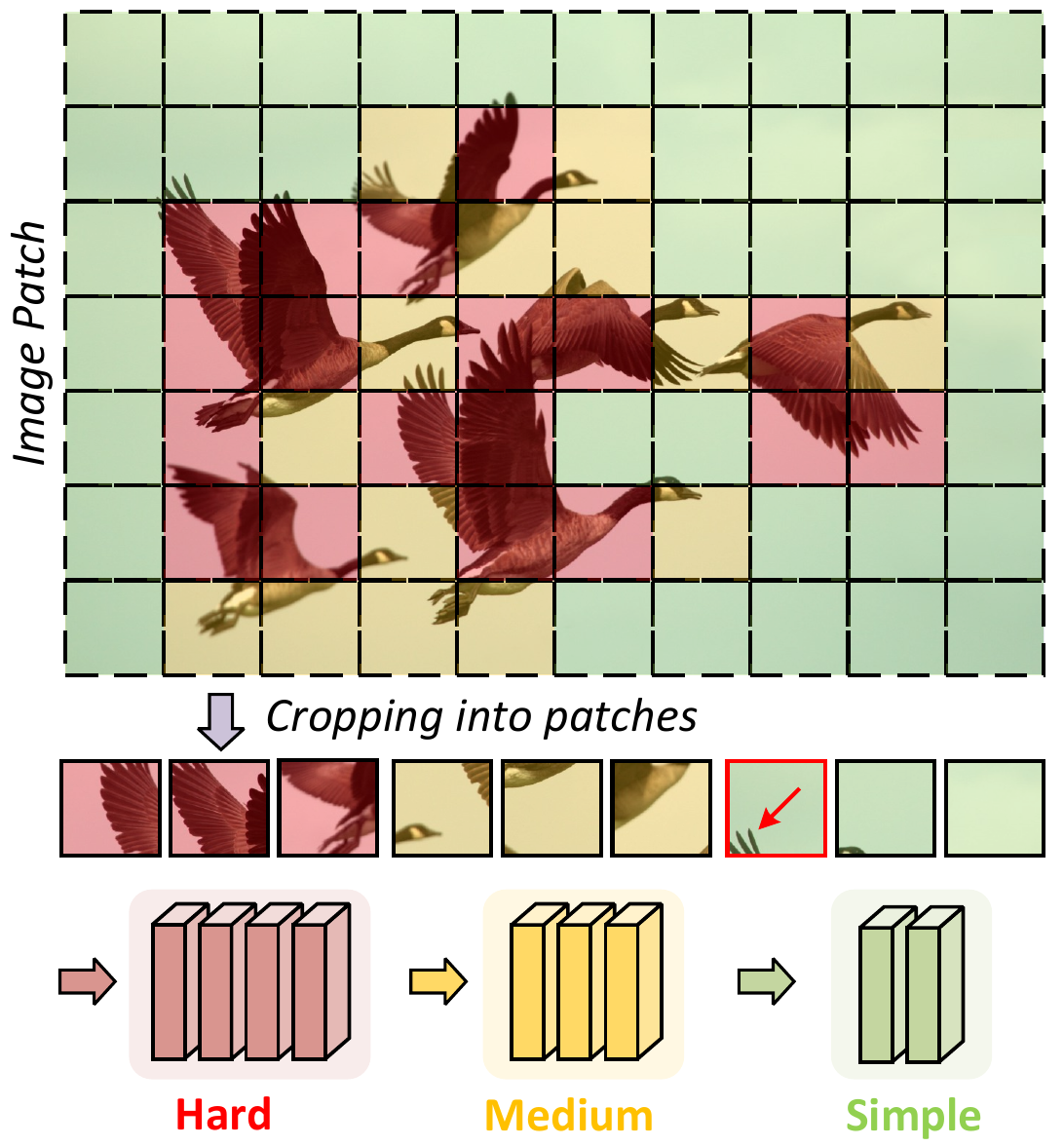} 
& \includegraphics[height=0.335\textwidth]{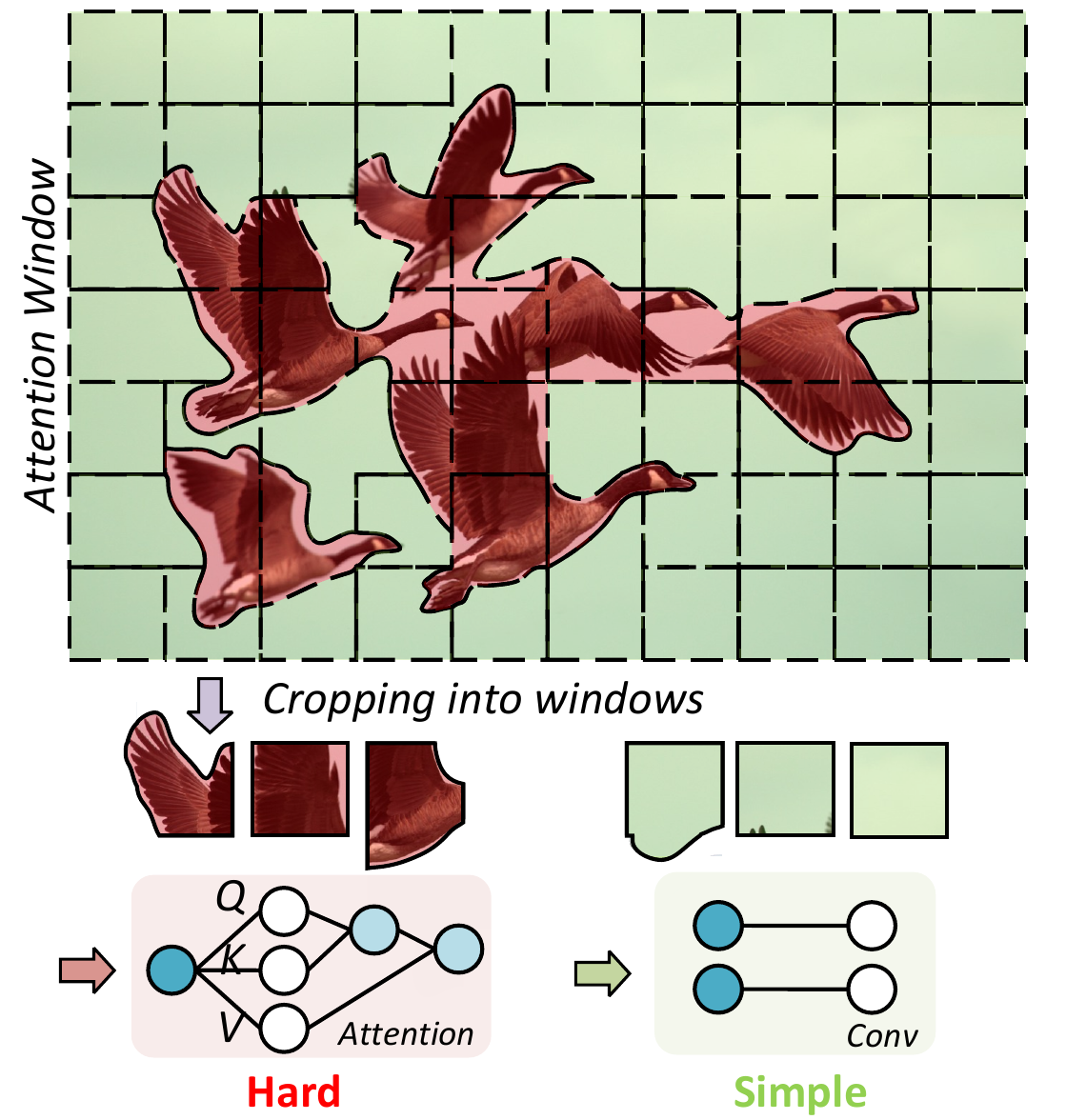} 
\end{tabular}
\captionof{figure}{{Comparison of ClassSR~\cite{ClassSR} framework and CAMixer.} Left) the plain/complex patches are at varied levels of difficulty to restore. Middle) ClassSR crops input images to sub-images for discriminative processing through models of varied complexities. Right) we introduce a content-aware mixer (CAMixer) to calculate self-attention for complexity regions while convolution for simple context.} 
\label{fig:start}
\end{center}}
\vspace{1em}
]

\setcounter{footnote}{0}
\renewcommand{\thefootnote}{\fnsymbol{footnote}}
\footnotetext[1]{Equal contribution. $^\dagger$Corresponding author.}
\footnotetext[3]{Work done during an internship at Bytedance. }
\renewcommand{\thefootnote}{}
\footnotetext{Code is available at: \url{www.github.com/icandle/CAMixerSR}}

\renewcommand{\thefootnote}{\arabic{footnote}}

\input{sec/0_abstract}    
\input{sec/1_intro}

\input{sec/2_formatting}
\input{sec/3_finalcopy}
{
    \small
    \bibliographystyle{ieeenat_fullname}
    \bibliography{main}
}

\end{document}

%% file: preamble.tex
\usepackage[dvipsnames]{xcolor}
\newcommand{\red}[1]{{\color{red}#1}}

\newcommand{\authorskip}{\hspace{5mm}}
\usepackage{listings}

\definecolor{tablegreen}{RGB}{11,225,83}
\definecolor{tablered}{RGB}{230,28,100}
\definecolor{tableblue}{RGB}{60,105,225}
\definecolor{tablegray}{gray}{0.45}

\newlength\savedwidth
\newcommand\whline{\noalign{\global\savedwidth\arrayrulewidth\global\arrayrulewidth 0.8pt}
\hline\noalign{\global\arrayrulewidth\savedwidth}}

\newcommand{\algname}{CAMixerSR}

\usepackage{soul}

\usepackage[ruled,linesnumbered]{algorithm2e}
\usepackage{amsmath,amsfonts,bm}

\def\eqref#1{equation~\ref{#1}}

\def\1{\bm{1}}

\def\rmA{{\mathbf{A}}}

\def\rmC{{\mathbf{C}}}

\def\rmF{{\mathbf{F}}}

\def\rmK{{\mathbf{K}}}

\def\rmM{{\mathbf{M}}}

\def\rmQ{{\mathbf{Q}}}

\def\rmV{{\mathbf{V}}}
\def\rmW{{\mathbf{W}}}
\def\rmX{{\mathbf{X}}}

\DeclareMathAlphabet{\mathsfit}{\encodingdefault}{\sfdefault}{m}{sl}
\SetMathAlphabet{\mathsfit}{bold}{\encodingdefault}{\sfdefault}{bx}{n}

\newcommand{\etens}[1]{\mathsfit{#1}}

\usepackage{multirow}
\usepackage[export]{adjustbox}
\usepackage{svg}

\definecolor{tablegreen}{RGB}{235,255,210}
\definecolor{tablered}{RGB}{255,255,220}
\definecolor{tableblue}{RGB}{210,235,255}

\definecolor{Highlight}{HTML}{39b54a}  %
\definecolor{red}{HTML}{ff3509} 
\definecolor{blue}{RGB}{15,45,245}
\definecolor{green}{RGB}{25,200,25}
\definecolor{lightgray}{gray}{0.9}%

\usepackage{appendix}
\usepackage{pifont}
\usepackage{xcolor}
\usepackage{colortbl}
\renewcommand{\red}[1]{\textbf{#1}}
\newcommand{\gray}[1]{\textcolor{gray}{#1}}
\newcommand{\blue}[1]{\underline{#1}}

\usepackage{arydshln}

%% file: sec/0_abstract.tex
\begin{abstract}
To satisfy the rapidly increasing demands on the large image (2K-8K) super-resolution (SR), prevailing methods follow two independent tracks: 1) accelerate existing networks by content-aware routing, and 2) design better super-resolution networks via token mixer refining. Despite directness, they encounter unavoidable defects (\eg, inflexible route or non-discriminative processing) limiting further improvements of quality-complexity trade-off. To erase the drawbacks, we integrate these schemes by proposing a content-aware mixer (CAMixer), which assigns convolution for simple contexts and additional deformable window-attention for sparse textures. Specifically, the CAMixer uses a learnable predictor to generate multiple bootstraps, including offsets for windows warping, a mask for classifying windows, and convolutional attentions for endowing convolution with the dynamic property, which modulates attention to include more useful textures self-adaptively and improves the representation capability of convolution. We further introduce a global classification loss to improve the accuracy of predictors. By simply stacking CAMixers, we obtain CAMixerSR which achieves superior performance on large-image SR, lightweight SR, and omnidirectional-image SR. 
\end{abstract}

%% file: sec/1_intro.tex
\section{Introduction}
\label{sec:intro}

Recent research on neural networks has significantly improved the image super-resolution (SR) quality~\cite{EDSR,RCAN,DDistillSR}. However, existing methods generate visual-pleasing high-resolution (HR) images but suffer intensive computations in real-world usages, especially for 2K-8K targets. To alleviate the overhead, many \textbf{accelerating frameworks}~\cite{ClassSR,ARM} and \textbf{lightweight networks}~\cite{IMDN,EFDN} were introduced for practical super-resolution application. 
However, these approaches are completely independent without cooperation. 

\begin{figure}
    \centering
    \includegraphics[width=0.95\linewidth]{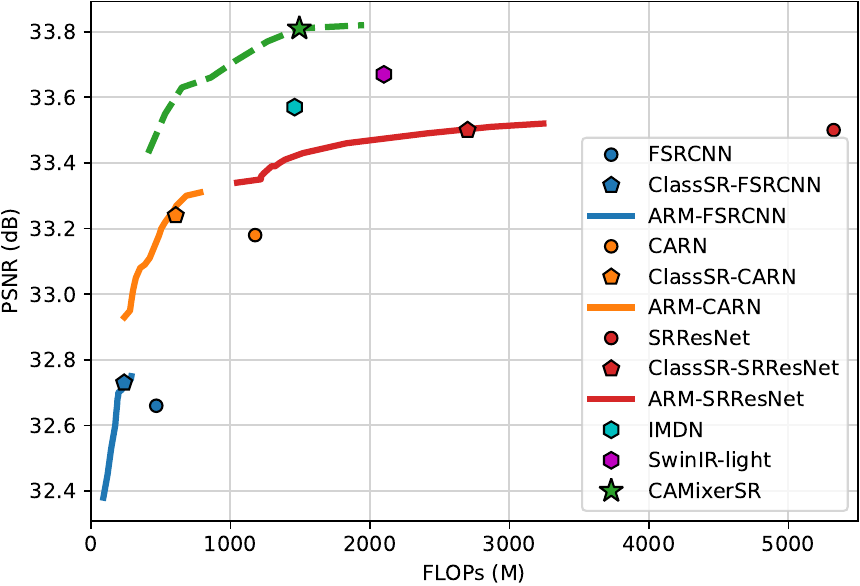}
    \caption{Performance (PSNR-FLOPs) comparison on Test8K. The green dashline indicates the trade-off curve of \algname{}.}
    \label{fig:classsr_cmp}
\end{figure}

The first strategy, the accelerating frameworks~\cite{ClassSR,PathRestore,hu2022restore}, is based on the observation that \emph{different image regions require different network complexities}, which tackles the problem from the perspective of content-aware routing of varied models. As depicted in the middle image of \cref{fig:start}, they decomposed large-input images into fixed patches and assigned patches to networks via an extra classification network.
ARM~\cite{ARM} further developed the strategy by introducing a LUT-based classifier and parameter-sharing design to improve efficiency. Despite these strategies being generic for all neural networks, two unavoidable defects remain. One is poor classification and inflexible partition. \cref{fig:start} displays the windows with little details that are improperly sent to a simple model. The other is the limited receptive fields. As shown in \cref{tab:crop}, cropping images into patches limits the receptive field, thus influencing the performance.

The second strategy, lightweight model design~\cite{SRCNN,VDSR,FMEN,PAN}, focuses on refining the neural operators (self-attention or convolution) and backbone structures to enable stronger feature representation capability within limited layers, \ie, using more intra-information to reconstruct images. For instance, NGswin~\cite{NGSwin} exploited N-Gram for self-attention to reduce the calculations and enlarge the receptive field. IMDN~\cite{IMDN} introduced information multi-distillation for efficient block design. Although these lightweight methods reached impressive efficiency on 720p/1080p images, their usages are rarely examined with larger input (2K-8K). Moreover, these approaches ignore intrinsic characteristics that different content can be discriminatively processed.

This paper, firstly integrating the above strategies, is based on the derived observation that \emph{distinct feature regions demand varying levels of token-mixer complexities}. As shown in~\cref{tab:class}, simple convolution (Conv) can perform similarly with much more complex convolution\,+\,self-attention (SA\,+\,Conv) for simple patches.
Hence, we propose a content-aware mixer (CAMixer) to route token mixers with different complexities according to the content. 
As depicted in~\cref{fig:start}, our CAMixer uses complex self-attention (SA) for intricate windows and simple convolution for plain windows. 
Furthermore, to address the limitations of ClassSR, we introduce a more sophisticated predictor. This predictor utilizes multiple conditions to generate additional valuable information, thereby enhancing CAMixer with improved partition accuracy and better representation.
Based on CAMixer, we construct \algname{} for super-resolution tasks. 
To fully examine the performance of CAMixer, we conduct experiments on both lightweight SR, large-input (2K-8K) SR, and omnidirectional-image SR. \cref{fig:classsr_cmp} illustrates \algname{} advances both lightweight SR and accelerating framework by a large margin. Our contributions are summarized as:
\begin{itemize}
    \item We propose a Content-Aware Mixer (CAMixer) integrating convolution and self-attention, which can adaptively control the inference computation by assigning simple areas to convolution and complex areas to self-attention.
    \item We propose a powerful predictor to generate the offset, mask, and simple spatial/channel attentions, which modulates CAMixer to capture longer-range correlation with fewer calculations.
    \item Based on CAMixer, we build \algname{} which exhibits state-of-the-art quality-computation trade-offs on three challenging super-resolution tasks: lightweight SR, large-input SR, and omnidirectional-image SR.
\end{itemize}

\begin{table}
  \centering
  \fontsize{8.5pt}{9.5pt}\selectfont
    \tabcolsep=10pt
\caption{PSNR (dB) values obtained by three token mixers.}
  \begin{tabular}{l|cccc}
    \whline
     {Method}
     & FLOPs & Simple & Medium & Hard \\
    \whline
    Conv & 517M & 43.73 & 30.96 & 23.60 \\
    SA + Conv & 979M & 43.80 & 31.19 & 23.80 \\
    CAMixer & 747M & 43.81 & 31.17 & 23.78 \\
    \whline
  \end{tabular}
  \label{tab:class}
\end{table}

\begin{table}
  \centering
  \fontsize{8.5pt}{9.5pt}\selectfont
    \tabcolsep=10pt
\caption{Cropping images into smaller tiles results in larger drops.}
  \begin{tabular}{l|ccc}
    \whline
     \multirow{2}{*}{Method} & \multicolumn{3}{c}{\{\emph{Tile, Overlap}\}}  \\
     & \{128,8\} & \{64,4\} & \{32,2\} \\
    \whline
    RCAN~\cite{RCAN} & 29.38 & 29.37 & 29.32 \\
    IMDN~\cite{IMDN} & 29.03 & 29.02 & 28.96 \\
    SwinIR-light~\cite{SwinIR}\quad\quad\quad & 29.24 & 29.22 & 29.17  \\
    \whline
  \end{tabular}
  \label{tab:crop}
\end{table}

\section{Related Work}
\noindent\textbf{Accelerating framework for SR}.
As the complexity continuously enlarged for better restoration quality, the practical application for SR models becomes harder, especially for 2K-8K SR. Recent research~\cite {ClassSR,PathRestore,ARM} tackled this problem from a different perspective. Instead of designing a lightweight model, they use content-aware routing to send cropped patches to models with different complexities dynamically. ClassSR~\cite{ClassSR} leveraged a 3-class classifier to determine the sub-image calculated by complex/medium/simple networks, which saves 50\% calculations for RCAN~\cite{RCAN} on 8K datasets. PathRestore~\cite{PathRestore} learned to select feature paths to adapt FLOPs according to context.  

\begin{figure*}
    \centering
    \includegraphics[width=1.0\textwidth]{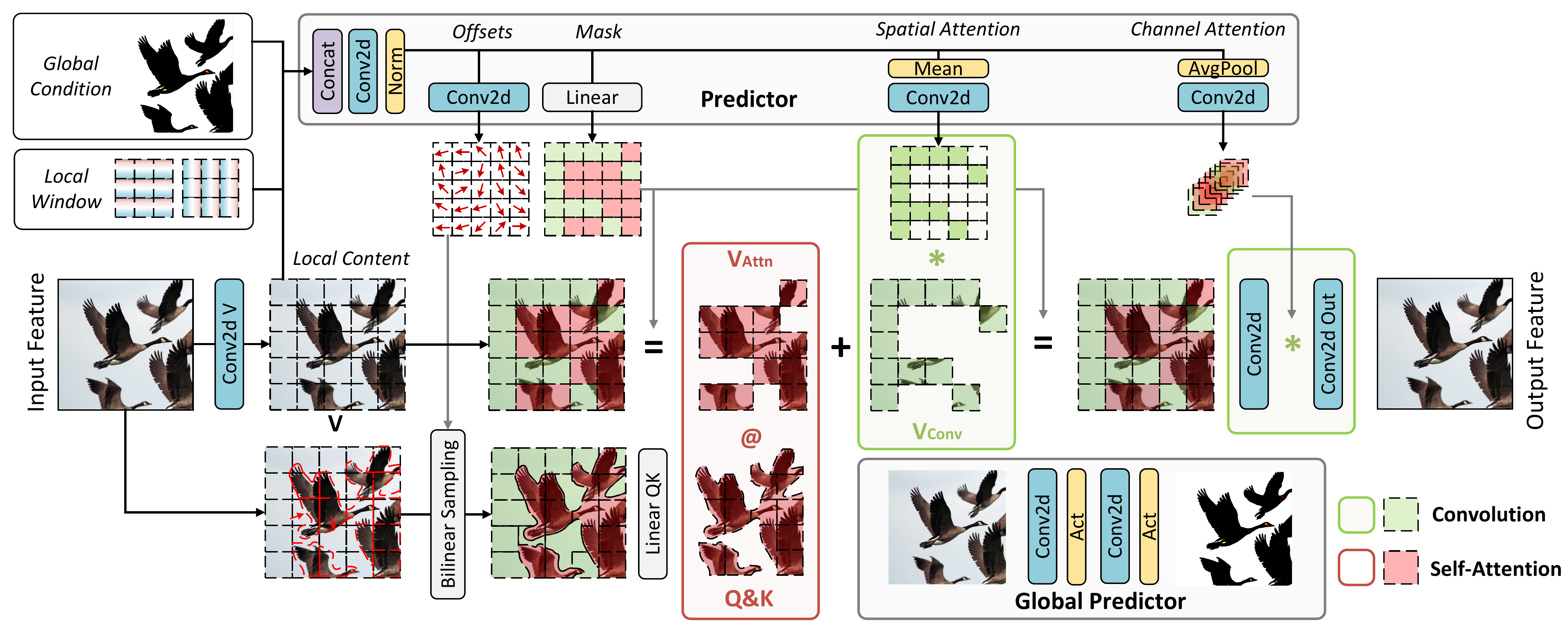}
    \caption{Overview of the proposed CAMixer. CAMixer consists of three parts: Predictor, Self-Attention branch, and Convolution branch.}
    \label{fig:CAMixer}
\end{figure*}

\noindent\textbf{Lightweight SR}.
Since the first work, SRCNN~\cite{SRCNN} using convolutional neural network (ConvNet) in the super-resolution task, numerous strategies~\cite{SRCNN,IMDN,RFDN,DDistillSR} to reduce the complexity have been proposed for a more lightweight inference. The early DRCN~\cite{DRCN} and DRRN~\cite{DRRN} tried to decrease parameters by using recurrent blocks but suffered intensive computations. To alleviate the drawback, IDN~\cite{IDN} and IMDN~\cite{IMDN} adopted efficient information fusion structure to reduce both parameters and calculations. The latter ConvNet, \eg, EFDN~\cite{EFDN} and RLFN~\cite{RLFN} further simplified the information distillation procedure and introduced reparameterization to obtain real-time inference on mobile devices.
With the recently rapid development of transformers, varied powerful token mixers, \eg, self-attention~\cite{SwinFIR,ELAN} and large kernel convolution~\cite{VapSR,MAN}, are introduced to lightweight SR. For example, SwinIR~\cite{SwinIR} utilized window-based SA, and MAN~\cite{MAN} employed large-kernel convolution, both of which achieved SOTA performance. Although these neural operators are capable of capturing long-range correlations, they cost massive calculations. To this end, this paper integrates the content-aware routing to token mixer design that adopts complex operators for informative areas but simple operators for plain areas.

%% file: sec/2_formatting.tex
\section{Method}

\subsection{Content-Aware Mixing}
The overview of our proposed CAMixer is shown in \cref{fig:CAMixer}. CAMixer consists of three main components: the predictor module, the attention branch, and the convolution branch.
Given the input features $\rmX\in\mathbb{R}^{C\times H\times W}$, it is first projected to attain \emph{value} $\rmV\in\mathbb{R}^{C\times H\times W}$ via a point-wise convolution:
\begin{equation}
    \rmV = f_{\etens{PWConv}}(\rmX),
    \label{eq:channel_aggr}
\end{equation}
\textbf{Predictor}. Based on the local condition $\rmC_l=\rmV$,  global condition $\rmC_{g}\in\mathbb{R}^{2\times H\times W}$, and linear positional encoding $\rmC_{w}\in\mathbb{R}^{2\times H\times W}$, the predictor first calculates the shared intermediate feature map $\rmF$ and then generate \textbf{offsets maps}, \textbf{mixer mask}, and simple \textbf{spatial/channel attention}:
\begin{equation}
\begin{aligned}
\rmF = f_{\etens{head}}(\rmC_l&, \rmC_{g},\rmC_{w}),\quad\hat{\rmF}= f_{\etens{reduce}}(\rmF)\in\mathbb{R}^{\frac{HW}{M^2} \times M^2},\\
\Delta p & = r\cdot f_{\etens{offsets}}(\rmF)\in\mathbb{R}^{2\times H\times W}, \\
  m &= \hat{\rmF}\rmW_{\etens{mask}}\in\mathbb{R}^{\frac{HW}{M^2}\times 1},\\ 
    \rmA_{s}& = f_{\etens{sa}}(\rmF)\in\mathbb{R}^{1\times H\times W}, \\
  \rmA_{c}& = f_{\etens{ca}}(\rmF)\in\mathbb{R}^{C\times 1\times 1}, \\
\end{aligned}
\end{equation}
where $\Delta p$ is the content-related offsets matrix to warp the window with more complex structures. $r$ is a scalar to control the offsets range. $\hat{\rmF}$ is the reduced and rearranged immediate feature according to the attention window size $M$. $m$ is the mask to decide whether the cropped window is calculated by attention or convolution. $\rmA_{s}$ and $\rmA_{c}$ are spatial and channel attention to enhance the convolutional branch.

\noindent\textbf{Attention Branch}. To calculate sparse attention for complex areas, we use the offsets $\Delta p$ to modulate the original input $\rmX$ by bilinear interpolation $\phi(\cdot)$ to include more useful content in selected windows:
\begin{equation}
    \Tilde{\rmX} = \phi(\rmX,\Delta p).
\end{equation}

We subsequently rearrange the $\Tilde{\rmX}$, ${\rmV}\in\mathbb{R}^{\frac{HW}{M^2}\times M^2\times C}$ according to the window shape $M\times M$. 
During the training stage, we apply the gumble softmax~\cite{jang2016categorical, DynamicViT} to calculate binary mask $\rmM = \etens{gumble\_softmax}(m)$ for hard and simple token sampling.
During inference, by descending sorting mask $m$ with $\etens{argsort}(m)$, we obtain the index $I_{\etens{hard}}$ of top $K$ related windows for sparse attention, and the $I_{\etens{simple}}$ of other $\frac{HW}{M^2}-K$ ones for convolution, where $K=\sum\rmM$. We denote the ratio of the attention patch as $\gamma=K/\frac{HW}{M^2}$. 
Upon the indices, we split the $\Tilde{\rmX}$ and ${\rmV}$ by:
\begin{equation}
\begin{aligned}
    \Tilde{\rmX}_{\etens{hard}} &= \Tilde{\rmX}[I_{\etens{hard}}]\in\mathbb{R}^{K\times M^2\times C},\\
    {\rmV}_{\etens{hard}} &= {\rmV}[I_{\etens{hard}}]\in\mathbb{R}^{K\times M^2\times C},\\
    {\rmV}_{\etens{simple}} &= {\rmV}[I_{\etens{simple}}]\in\mathbb{R}^{(\frac{HW}{M^2}-K)\times M^2\times C}.
\end{aligned}
\end{equation}

After obtaining $\Tilde{\rmX}_{\etens{hard}}$, the \emph{query} $\Tilde{\rmQ}_{\etens{hard}}$ and \emph{key} $\Tilde{\rmK}_{\etens{hard}}$ are generated by linear layers:
\begin{equation}
    \Tilde{\rmQ} = \Tilde{\rmX}_{\etens{hard}}\rmW_q, \quad \Tilde{\rmK} = \Tilde{\rmX}_{\etens{hard}}\rmW_k\in\mathbb{R}^{K\times M^2\times C}.
\end{equation}

Based on the above deduction, the self-attention for complex windows can be expressed as:
\begin{equation}
    {\rmV}_{\etens{hard}} = \etens{softmax}(\frac{\Tilde{\rmQ}\Tilde{\rmK}^{T}}{\sqrt{d}}){\rmV}_{\etens{hard}}\in\mathbb{R}^{K\times M^2\times C}.
\end{equation}

For windows ${\rmV}_{\etens{simple}}$ for light operation, we use the rearranged $\rmA_{s}$ to implement the simple attention by element-wise multiplication: 
\begin{equation}
    {\rmV}_{\etens{simple}} = {\rmA}_{s}\cdot{\rmV}_{\etens{simple}}\in\mathbb{R}^{(\frac{HW}{M^2}-K)\times M^2\times C}.
\end{equation}

Overall, we integrate the ${\rmV}_{\etens{hard}}$ and ${\rmV}_{\etens{simple}}$ to obtain the output of attention branch ${\rmV}_{\etens{attn}}\in\mathbb{R}^{C\times H\times W}$ with the help of the the indices.

\noindent\textbf{Convolution Branch}. 
We leverage a depth-wise convolution and pre-generated channel attention to capture the local correlation, which can be formulated as:
\begin{equation}
    {\rmV}_{\etens{conv}} = f_{\etens{DWConv}}({\rmV}_{\etens{attn}})\cdot\rmA_c + {\rmV}_{\etens{attn}}
\end{equation}

Finally, the output of CAMixer is projected by a point-wise convolution as:
\begin{equation}
    \rmV_{\etens{out}} = f_{\etens{PWConv}}({\rmV}_{\etens{conv}}),
    \label{eq:channel_aggr0}
\end{equation}

\begin{table}
  \centering
  \tabcolsep=2pt
\fontsize{8.5pt}{9.5pt}\selectfont
  \caption{Complexity comparison between convolution, window-based multi-head self-attention, and our CAMixer. CAMixer semantically integrates convolution and self-attention with changeable calculations. $h, w$: height and width of the input image. $C$: input and output channel. $k$: convolution kernel size. $M$: window size. $\gamma$: the ratio of tokens calculated by SA. $\rho$: the reduction ratio.}
  \label{tab:complexity}
  \begin{tabular}{lcc@{}}
    \whline
     Method & FLOPs ($\times hw$) & Params \\
    \whline
    Conv & $k^2C^2$ & $k^2C^2$ \\
    \hline
    W-MSA & $4C^2 + 2M^2C$ & $4C^2$\\
    \hline
      \multirow{3}{*}{CAMixer}  & $k^2C$ +  & $k^2C$\\
     & $2(1+\gamma)C^2 + 2\gamma M^2C$ + & $4C^2$\\
     & $\underbrace{\rho C(C+4)}_{\etens{shared}} + \underbrace{M + 2\rho C}_{\etens{mask\&offsets}} + \underbrace{\rho K^2C + {\rho C^2}/{hw}}_{\etens{spatial\&channel}}$ & $<3\rho C^2$ \\
    \whline
  \end{tabular}
\end{table}

\noindent\textbf{CAMixer}. Overall, by controlling the self-attention ratio $\gamma$, we adjust the content-aware mixing. When $\gamma=1$, CAMixer is a combination of self-attention and convolution, which is similar to ACMix~\cite{ACMix}. For $\gamma=0$, CAMixer is a pure convolutional token mixer with  low complexity.
For $\gamma\in(0,1)$, CAMixer learns the content-aware mixing that uses complex mode for hard areas but simple mode for plain areas.

\noindent\textbf{Complexity Analysis}. We theoretically compare the complexity of convolution, window-based self-attention, and CAMixer in \cref{tab:complexity}. Specifically, the FLOPs of CAMixer consist of three parts: the convolution, the attention, and the predictor. Given the input of $C\times h\times w$, the convolution branch utilizes a depth-wise convolution with a computational cost $K^2Chw$. For attention branch, four projection operations cost $2(1+\gamma)C^2hw$ while the attention calculation costing $2\gamma M^2Chw$, where $\gamma={K}/(\frac{hw}{M^2})$ is the ratio of hard windows. For predictor module, it adds a serial of computations: $\rho C(C+4)hw$ for shared head, $Mhw$ for mask, $2\rho Chw$ for offsets, $\rho k^2Chw$ for spatial attention, and $\rho C^2$ for channel attention, where $\rho=\frac{1}{8}$ is the reduction ratio to reduce calculation.

\subsection{Network Architecture}
Lastly, we construct the \algname{} by modifying SwinIR-light~\cite{SwinIR}. Generally, \algname{} consists of four components, three from SwinIR: shallow feature extractor, deep feature extractor, reconstruction module, and additional global predictor module shown in~\cref{fig:CAMixer}. Furthermore, we replaced the window-based self-attention with CAMixer and reduced the block number.

\subsection{Training Loss}
We describe the training objectives of our \algname{}, including the optimization of the super-resolution framework and the predictor.
Following the previous work~\cite{EDSR,DDistillSR}, we adopt the primary $\ell_1$ loss to train the backbone.
Assuming the input batch with $N$ image pairs, \ie, $\{I_i^{\etens{LR}},I_i^{\etens{HR}}\}_{i=1}^{N}$, this process can be formulated by:
\begin{equation}
    \ell_1 = \frac{1}{N}\sum_{i=1}^{N}\left\| I_i^{\etens{HR}}-f_{\etens{\algname{}}}(I_i^{\etens{LR}}) \right\|_1,
\end{equation}
where $f_{\etens{\algname{}}}(\cdot)$ is the proposed \algname{}.

To supervise the predictors for $S$ CAMixers, following previous works that control SA sparsity~\cite{DynamicViT, EvoViT}, we adopt a simple but effective MSE Loss to control the ratio $\gamma_i$:
\begin{equation}
    \ell_{\etens{ratio}} = \frac{1}{N}\sum_{i=1}^{N}\left\| \gamma_\etens{ref} \cdot\left(1- \frac{2}{S}\sum_{i=1}^{S}\gamma_i\right) \right\|_2,
\end{equation}
where $\gamma_\etens{ref}$ represents the referred overall ratio and $\gamma_i$ denotes the hard token ratio for $i$-th CAMixer. To enable one training for dynamic ratio, we pre-train the \algname{} with $\gamma_\etens{ref}=0.5$ and fine-tune with $\gamma_\etens{ref}\in [0, 1]$. 

Overall, we train the \algname{} by simply combining the above objectives:
\begin{equation}
    \ell = \ell_1 + \ell_\etens{ratio}.
\label{eq:loss}
\end{equation}

\begin{figure}
    \centering
      \fontsize{8.5pt}{9.5pt}\selectfont
    \tabcolsep=1pt
    \begin{tabular}{cccc}

       \rotatebox{90}{\quad$\rmC_w$\quad$\rmC_g$\quad\,\,$\rmC_l$} &\includegraphics[width=0.12\linewidth,height=0.28\linewidth]{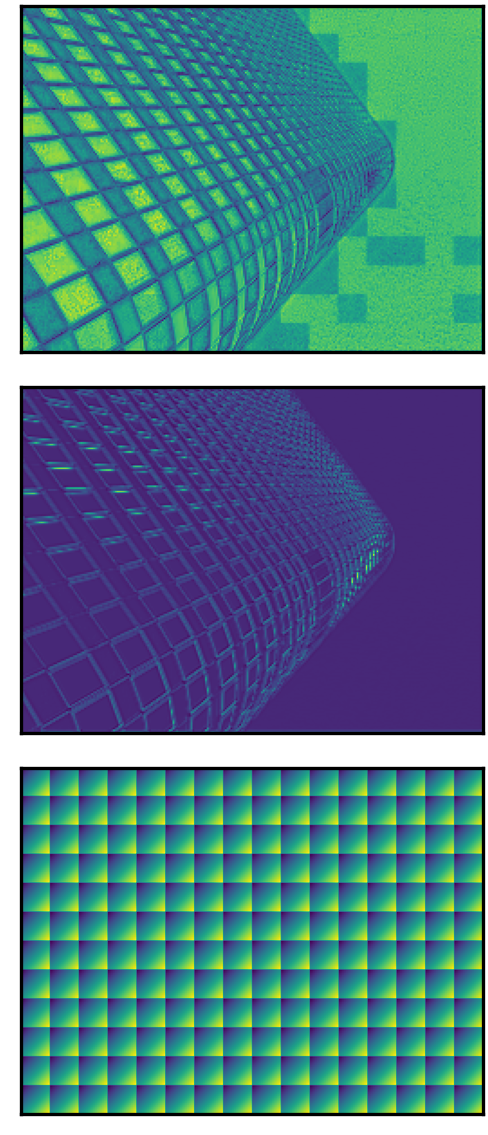} 
        & \includegraphics[width=0.4\linewidth,height=0.28\linewidth]{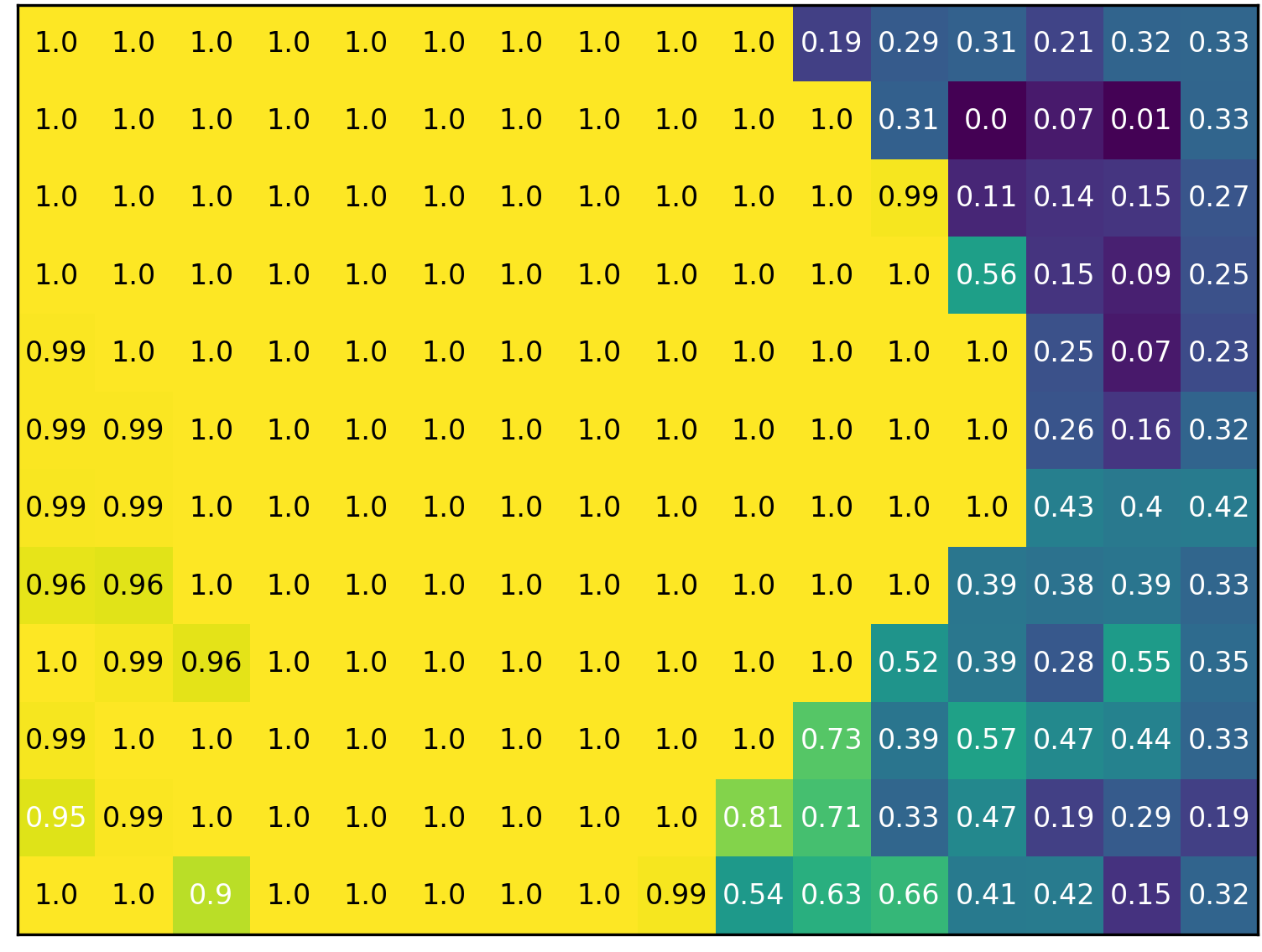}
        & \includegraphics[width=0.4\linewidth,height=0.28\linewidth]{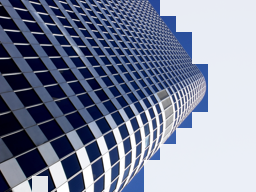}
        \\
        \rotatebox{90}{\quad$\rmC_w$\quad$\rmC_g$\quad\,\,$\rmC_l$} 
        &\includegraphics[width=0.12\linewidth,height=0.28\linewidth]{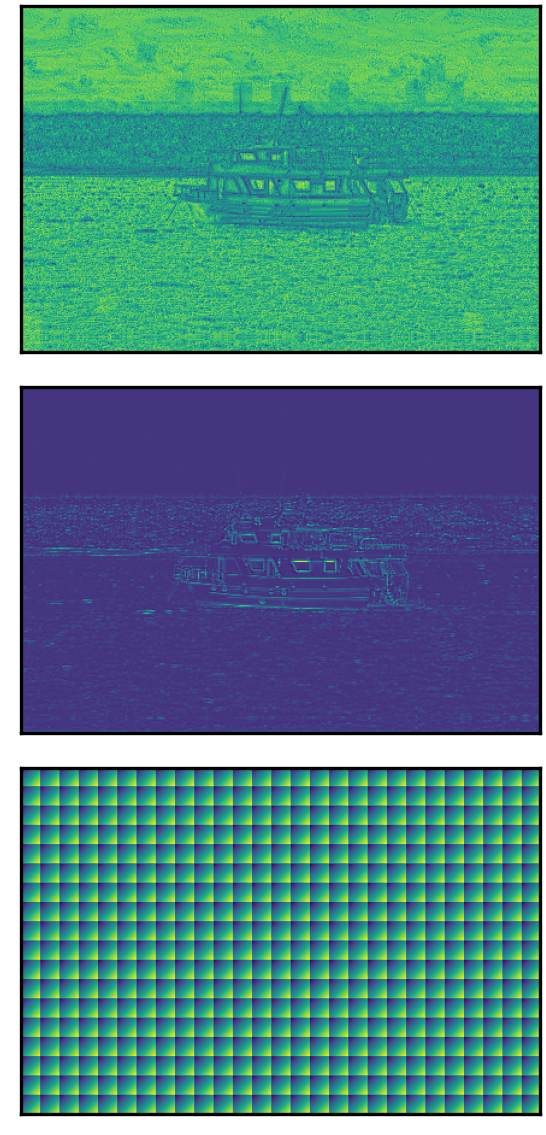} 
        & \includegraphics[width=0.4\linewidth,height=0.28\linewidth]{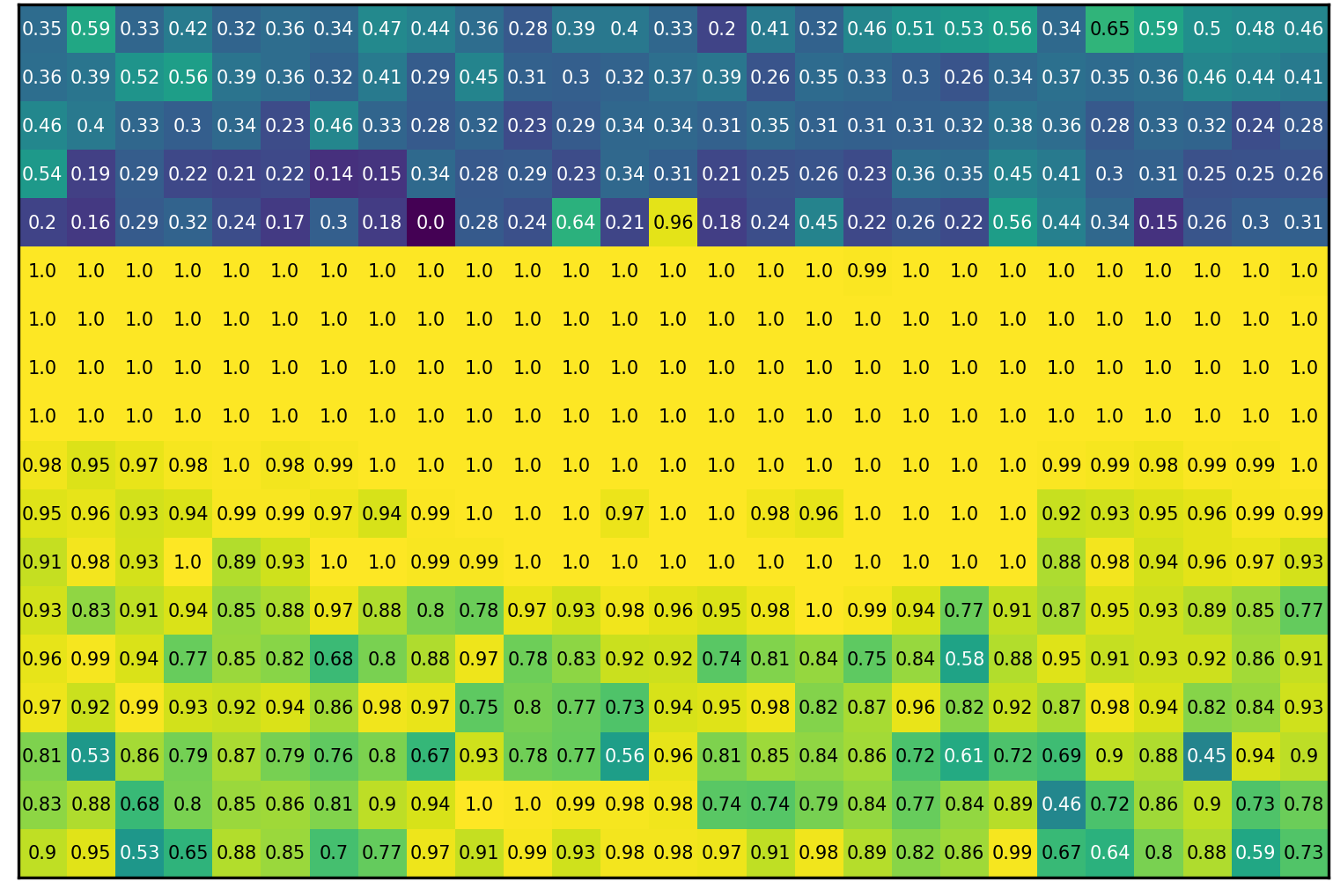}
        & \includegraphics[width=0.4\linewidth,height=0.28\linewidth]{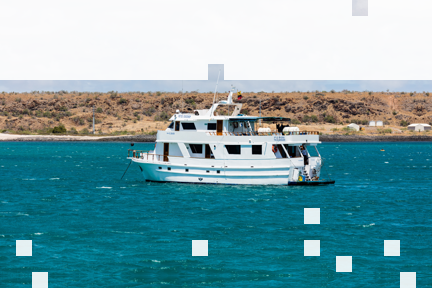}
        \\
        \rotatebox{90}{\quad$\rmC_w$\quad$\rmC_g$\quad\,\,$\rmC_l$} 
        &\includegraphics[width=0.12\linewidth,height=0.28\linewidth]{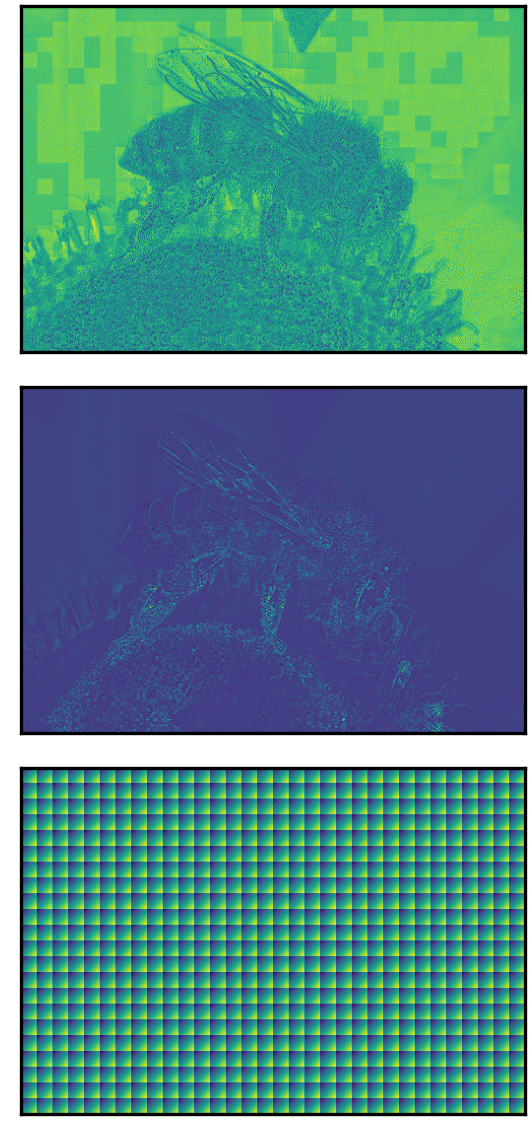} 
        & \includegraphics[width=0.4\linewidth,height=0.28\linewidth]{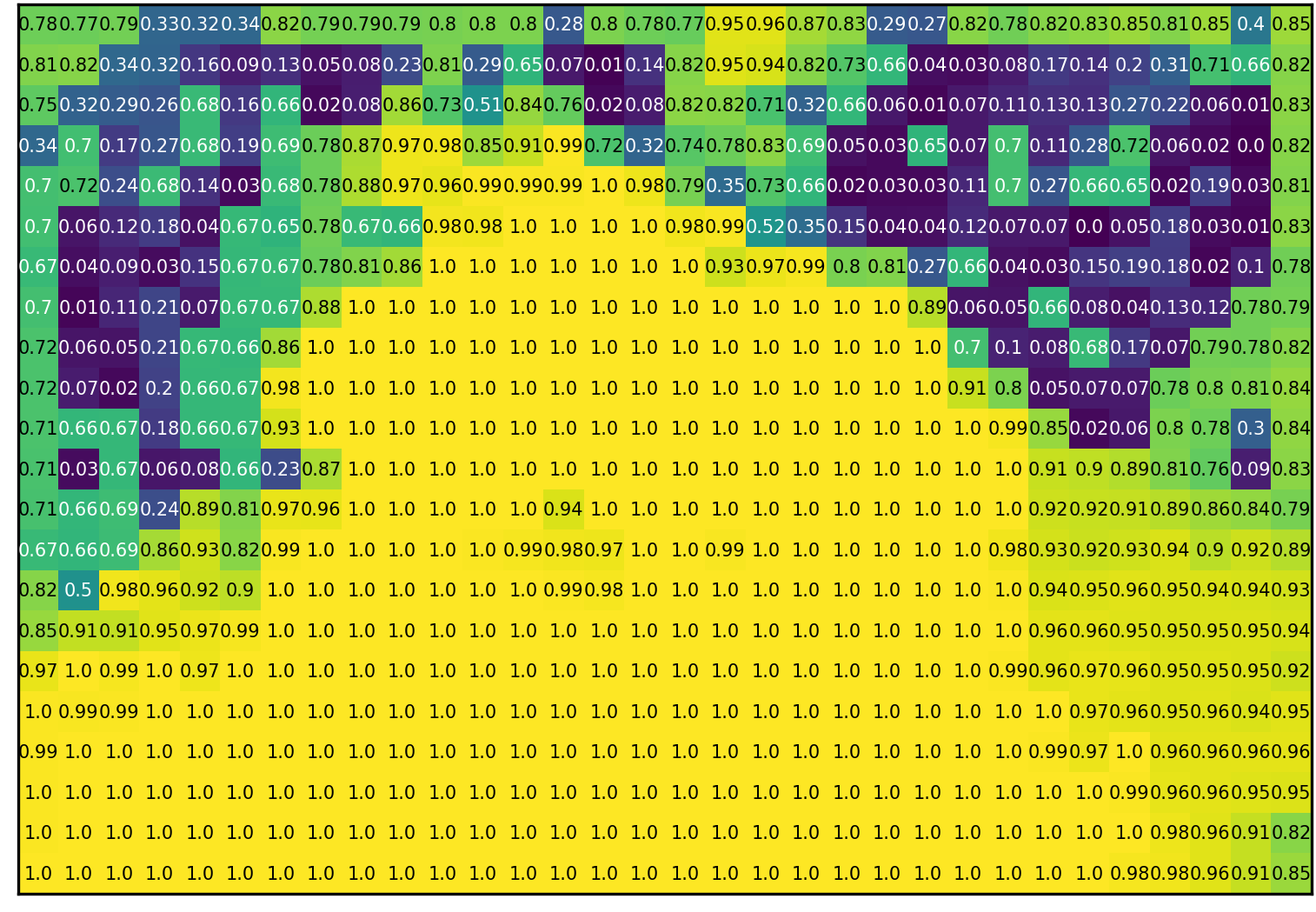}
        & \includegraphics[width=0.4\linewidth,height=0.28\linewidth]{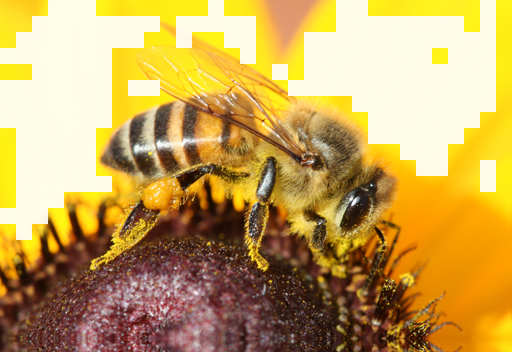}\\
        &Input & Predicted Mask $m$ & Masked Image
    \end{tabular}
    \caption{Visualizations of predicted mixer mask $m$ of \algname{}. The lighter the color, the larger the magnitude. The scores of attention windows are in black, and the ones of convolution are in white. The unmasked tokens with more complex content (higher score) are processed by self-attention.}
    \label{fig:mask}
\end{figure}

\section{Experiment}
\subsection{Implementation Details}
\noindent\textbf{Model}. Following SwinIR~\cite{SwinIR} and ELAN~\cite{ELAN}, we construct the overall backbone with 20 CAMixer and FFN blocks. The channel number is 60. The window size of self-attention is 16 and the convolution branch is implemented by two 3$\times$3 depth-wise convolution.  {Specifically, we manually set the $\gamma=1.0$ as the {\emph{Original}} model (baseline) and $\gamma=0.5$\footnote{For model, $\gamma$ is the averaged ratio for all CAMixer.} as the {\emph{CAMixer}} model} (target). We provide more results of other settings in \emph{supplementary material}. 

\noindent\textbf{Training}.
We train the proposed framework on three challenging super-resolution (SR) tasks: \textbf{\emph{lightweight SR}}, \textbf{\emph{large-image SR}}, and \textbf{\emph{omnidirectional-image (ODI) SR}}. For the first two tasks, we use DIV2K~\cite{div2k} as the training set. For ODISR, we leverage the cleaned ODI-SR dataset~\cite{LAU-Net}. The loss in \cref{eq:loss} is calculated with batch-size 32 and patch-size 64. The AdamW~\cite{AdamW} is adopted with the initial learning rate 5$\times 10^{-4}$ and 500$k$ iterations training procedure. We halved the learning rate at 250$k$, 400$k$, 450$k$, and 475$k$.

\noindent\textbf{Testing}. We first test {\algname{}-\emph{Original}} with $\gamma=1.0$ as the baseline, which represents all tokens being processed by self-attention and convolution. Then, we validate the proposed {\algname{}} with $\gamma=0.5$ which uses self-attention for partial tokens. For \textbf{\emph{lightweight SR}}, we employ five common-used validation datasets: Set5~\cite{Set5}, Set14~\cite{Set14}, BSD100~\cite{B100}, Urban100~\cite{Urban100}, and Manga109~\cite{manga109}. For \textbf{\emph{large-image SR}}, we utilize Flickr2K~\cite{EDSR} (F2K) and DIV8K~\cite{div8k} (Test2K, Test4K, and Test8K) to generate the testing datasets as ClassSR. For \textbf{\emph{ODI SR}}, we evaluate our models on the ODI-SR~\cite{LAU-Net} testing set and SUN360~\cite{sun360} dataset. For evaluation, we use PSNR and SSIM~\cite{SSIM} and additionally distortion-weighted versions WS-PSNR~\cite{WS-PSNR} and WS-SSIM~\cite{WS-SSIM} for ODI-SR.

\begin{table}
  \centering
  \fontsize{8.5pt}{9.5pt}\selectfont
    \tabcolsep=2pt
    \caption{Ablation study on window size on Urban100.}
  \begin{tabular}{c|c|cc|cc}
    \whline
    \multirow{2}{*}{Window Size} & \multirow{2}{*}{\#Params}  & \multicolumn{2}{c|}{\emph{Original}} & \multicolumn{2}{c}{\emph{CAMixer}} \\
    & & \#MAdds & PSNR & \#MAdds & PSNR\\
    \whline
    8$\times$8 & 693K & 47.5G & 26.42  & 38.6G \textcolor{gray}{(81.3\%)} & 26.41\\
    16$\times$16 & 765K &  77.9G & 26.65 & 53.8G \textcolor{gray}{(69.1\%)} & 26.63\\
    32$\times$32 & 1340K & 191.2G & 26.82 & 109.7G \textcolor{gray}{(57.4\%)} & 26.74\\
    \whline
  \end{tabular}
  \label{tab:windows}
\end{table}

\begin{figure}
\centering
\includegraphics[height=0.39\linewidth]{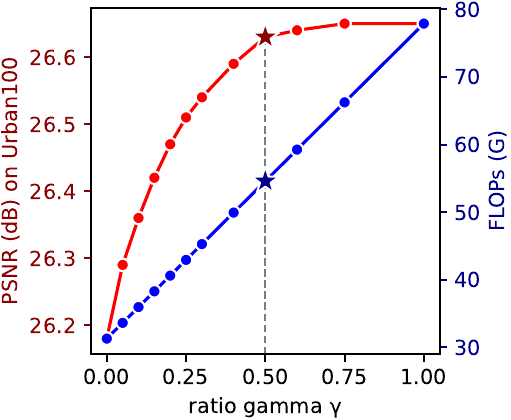}
\includegraphics[height=0.39\linewidth]{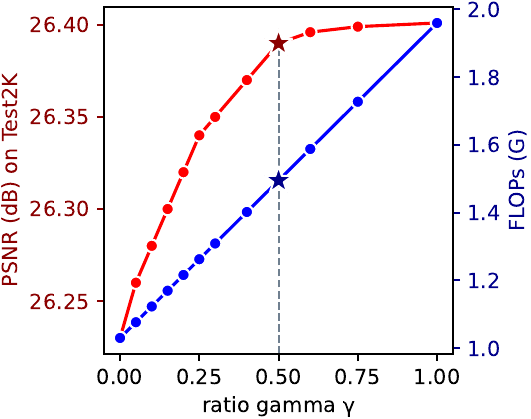}
\caption{Ablation study on attention ratio $\gamma$. }
\label{fig:gamma}
\vspace{-2mm}
\end{figure}

\begin{figure*}
    \centering      
    \fontsize{8.5pt}{9.5pt}\selectfont
    \tabcolsep=1pt
    \begin{tabular}{ccccccccc}
    & Block 1 & Block 4 & Block 7 & Block 10 & Block 14 &  Block 17 & Block 20 \\
         \rotatebox{90}{$\gamma=0.75$} & \includegraphics[width=0.13\linewidth]{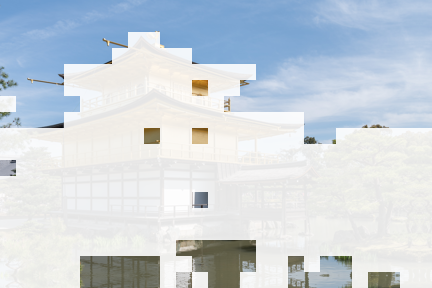}
         &  \includegraphics[width=0.13\linewidth]{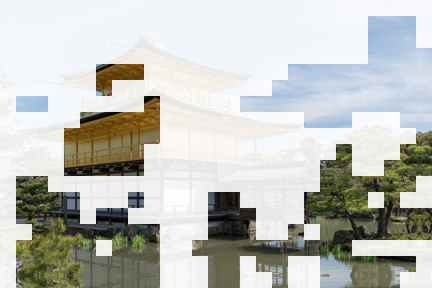}
         &  \includegraphics[width=0.13\linewidth]{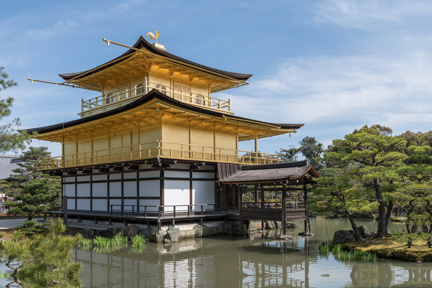}
         &  \includegraphics[width=0.13\linewidth]{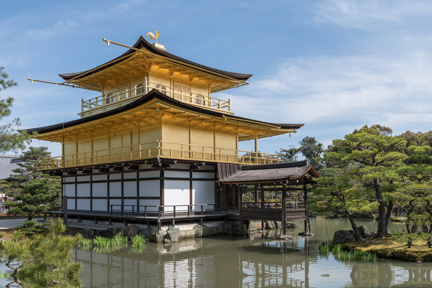}
         &  \includegraphics[width=0.13\linewidth]{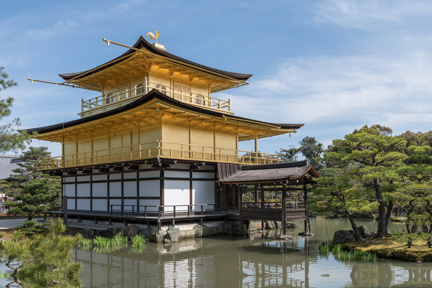}
         &  \includegraphics[width=0.13\linewidth]{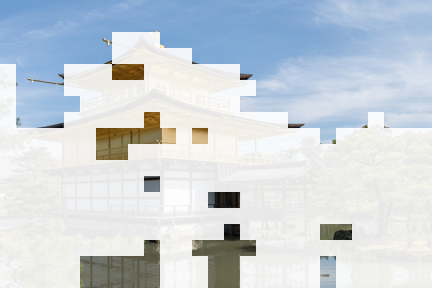}
         & \includegraphics[width=0.13\linewidth]{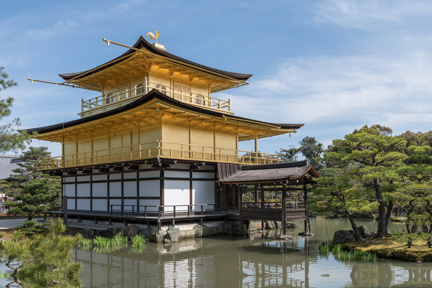}
         & \rotatebox{90}{$\gamma'=0.705$} \\
         \rotatebox{90}{$\gamma=0.50$} & \includegraphics[width=0.13\linewidth]{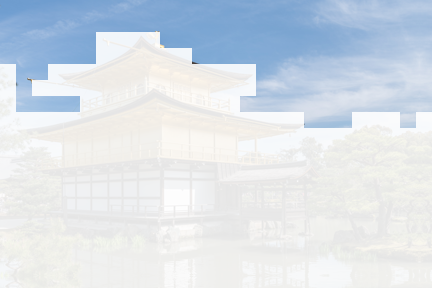}
         &  \includegraphics[width=0.13\linewidth]{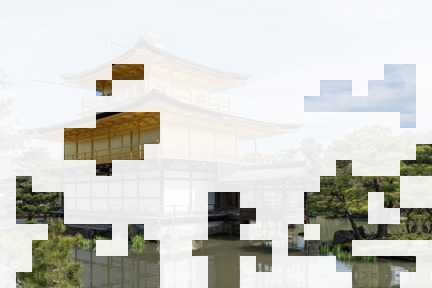}
         &  \includegraphics[width=0.13\linewidth]{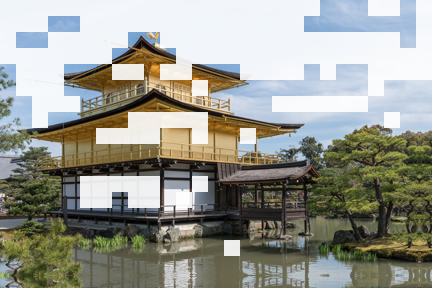}
         &  \includegraphics[width=0.13\linewidth]{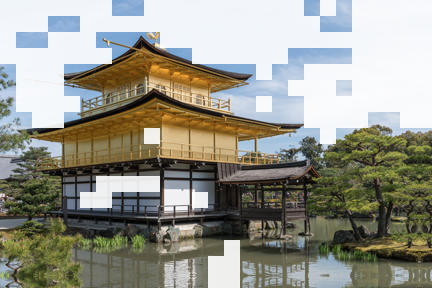}
         &  \includegraphics[width=0.13\linewidth]{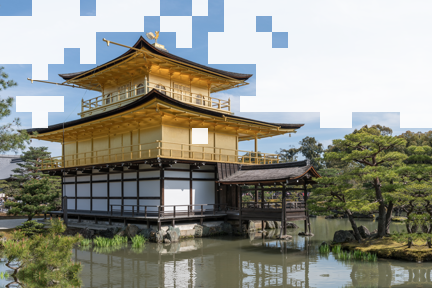}
         &  \includegraphics[width=0.13\linewidth]{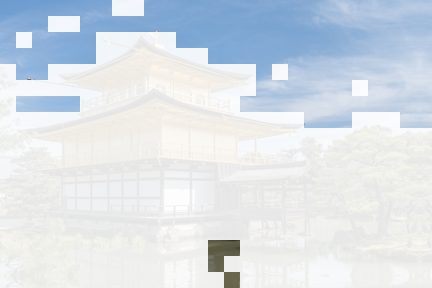}
         & \includegraphics[width=0.13\linewidth]{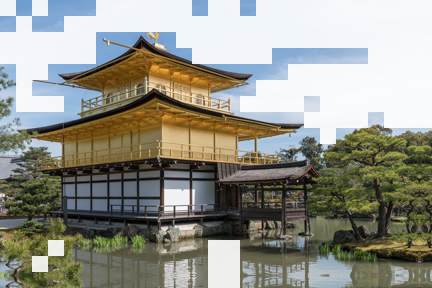}
         & \rotatebox{90}{$\gamma'=0.508$} \\
        \rotatebox{90}{$\gamma=0.25$} & \includegraphics[width=0.13\linewidth]{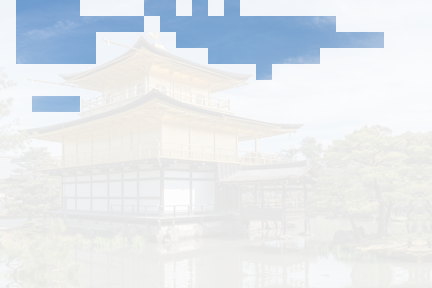}
         &  \includegraphics[width=0.13\linewidth]{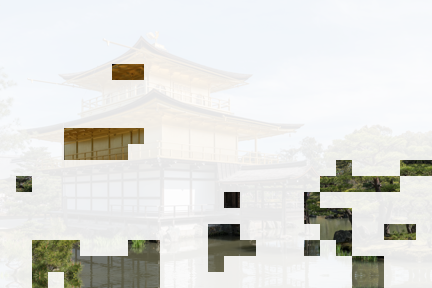}
         &  \includegraphics[width=0.13\linewidth]{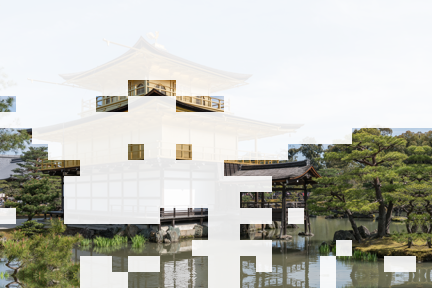}
         &  \includegraphics[width=0.13\linewidth]{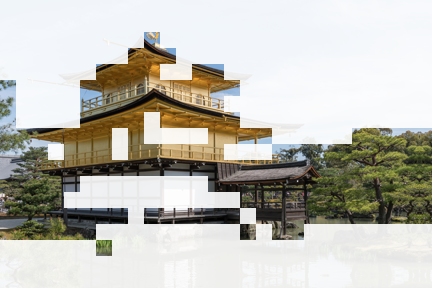}
         &  \includegraphics[width=0.13\linewidth]{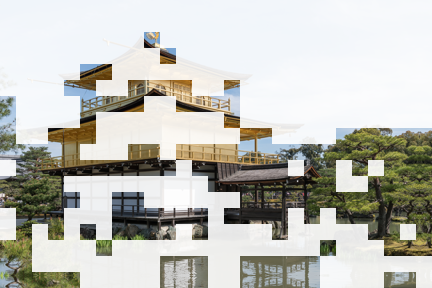}
         &  \includegraphics[width=0.13\linewidth]{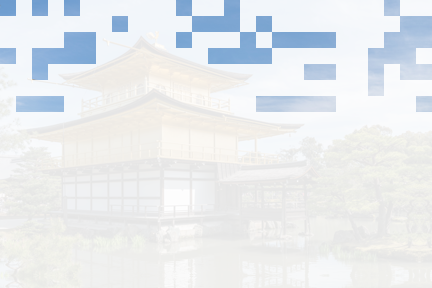}
         &  \includegraphics[width=0.13\linewidth]{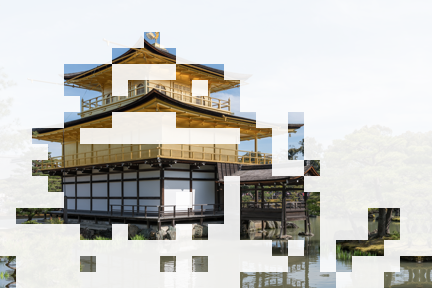}
         & \rotatebox{90}{$\gamma'=0.247$} \\
         $K$ & 199$\rightarrow$143$\rightarrow$62 & 193$\rightarrow$126$\rightarrow$61 & 486$\rightarrow$345$\rightarrow$177 & 486$\rightarrow$343$\rightarrow$177 & 486$\rightarrow$359$\rightarrow$179 & 195$\rightarrow$138$\rightarrow$52 & 486$\rightarrow$351$\rightarrow$162 \\
    \end{tabular}
    \caption{Visualization of the progressively classified tokens under varied ratio $\gamma$. The unmasked areas are processed by self-attention while the masked ones are processed by convolutional-attention. $K=\sum \rmM$ is the window number of the SA branch. $\gamma$ and $\gamma'$ are the target and actual ratio of SA tokens, respectively. }
    \label{fig:gamma0}
    \vspace{-3mm}
\end{figure*}

\subsection{Ablation Study}

\noindent\textbf{Effects of Self-Attention}.
We first compare the window size $M$ for the SA branch in~\cref{tab:windows}. The model with window size 32 is about 2.5$\times$ larger than window 16 while improving 0.17dB on Urban100. The model with window 16 advances window 8 by 0.23dB but only increases 16G MAdds. Furthermore, when we set $\gamma=0.5$, the 32$\times$32 window encounters larger performance drops than the other two models because a large window is hard to classify. To this end, we use window 16$\times$16 for better trade-offs between performance and computations. Moreover, we compare the different self-attention ratios $\gamma$ in~\cref{fig:gamma}. For both lightweight SR and 2K SR, the computations grow linearly, while the PSNR grows faster when $\gamma<0.5$ but slower when $\gamma>0.5$. Hence, we manually select $\gamma=0.5$ where PSNR is almost the same as $\gamma=1.0$ but reduce half the computations of SA.

\noindent\textbf{Effects of predictor components}. Compared to the simple classifiers~\cite{ClassSR,ARM} outputting only decision score, the proposed predictor generates more useful metrics (offsets and spatial/channel attention) for better partition and representation. We examine these extra components in \cref{tab:Predictor}. Removing any of them would result in huge performance drops. In detail, the offsets bring about 0.05dB gains and convolutional attentions jointly obtain 0.1dB improvements.

\noindent\textbf{Effects of mixer mask $m$}. To study the preference of the mixer mask dividing the hard and simple tokens, we visualize the learned mask in \cref{fig:mask}. The proposed predictor can learn proper masks to assign complex tokens (\eg, building, ship, and bee) to the attention branch while plain tokens (\eg, sky and petal) to the convolution branch. In \cref{fig:gamma0}, we further compare the partition masks for varied blocks under different ratios $\gamma$. The token number $K$ for SA is decreased according to the $\gamma$ and the actual ratio is close to the target. This significantly satisfies our goal to control the inference complexity. Most blocks learn to use SA for complex content. Interestingly, they focus on semantically different contexts, \eg, \emph{Block 7} preserving planets, \emph{Block 14} preserving edges, and \emph{Block 20} preserving buildings. Moreover, in \cref{fig:mask}, the $\rmC_l$ illustrates that the output of a convolution is distinctive from the attention. Thus, for \emph{Block 1\&17}, they employ SA for plain windows to harmonize the feature.

\begin{table}
  \centering
  \fontsize{8.5pt}{9.5pt}\selectfont
    \tabcolsep=5pt
  \caption{Ablation study on predictor components.}
  \begin{tabular}{ccc|c|cc}
    \whline
    Offsets $\Delta p$ & SA $\rmA_{s}$ &  CA $\rmA_{c}$ & \#Param & Set5 & Urban100 \\
    \whline
    \ding{51} & \ding{51} & \ding{51}
    & 765K & \textbf{32.51} & \textbf{26.63}\\
    \ding{51} & \ding{51} & {\color{gray}\ding{55}}
    & 745K & 32.43 & 26.50\\
    \ding{51} & {\color{gray}\ding{55}} & \ding{51}
    & 762K & 32.37 & 26.46\\
    {\color{gray}\ding{55}} & \ding{51} & \ding{51}
    & 762K & 32.46 & 26.60\\
    \ding{51} & {\color{gray}\ding{55}} & {\color{gray}\ding{55}} 
    & 742K & 32.31 & 26.45\\
    \whline
  \end{tabular}
  \label{tab:Predictor}
\end{table}

\begin{table}
  \centering
  \fontsize{8.5pt}{9.5pt}\selectfont
    \tabcolsep=6pt
  \caption{Ablation study on input conditions.}
  \begin{tabular}{ccc|cc}
    \whline
    Local $\rmC_l$ & Global $\rmC_{g}$ &  Window $\rmC_{w}$  & Set5 & Urban100 \\
    \whline
    \ding{51} & \ding{51} & \ding{51}
    & \textbf{32.51} & \textbf{26.63}\\
    \ding{51} & \ding{51} & {\color{gray}\ding{55}}
     & 32.49 & 26.55\\
    \ding{51} & {\color{gray}\ding{55}} & \ding{51}
    & 32.45 & 26.54\\
    \ding{51} & {\color{gray}\ding{55}} & {\color{gray}\ding{55}} 
    & 32.42 & 26.46\\
    \whline
  \end{tabular}
  \label{tab:condition}
\end{table}

\begin{figure}
    \centering
      \fontsize{8.5pt}{9.5pt}\selectfont
    \tabcolsep=1pt
\begin{tabular}{cccc}
     \includegraphics[width=0.24\linewidth]{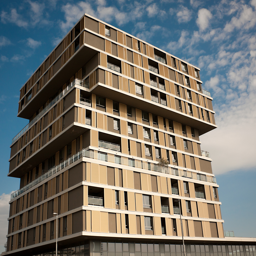}
     & \includegraphics[width=0.24\linewidth,cframe=black]{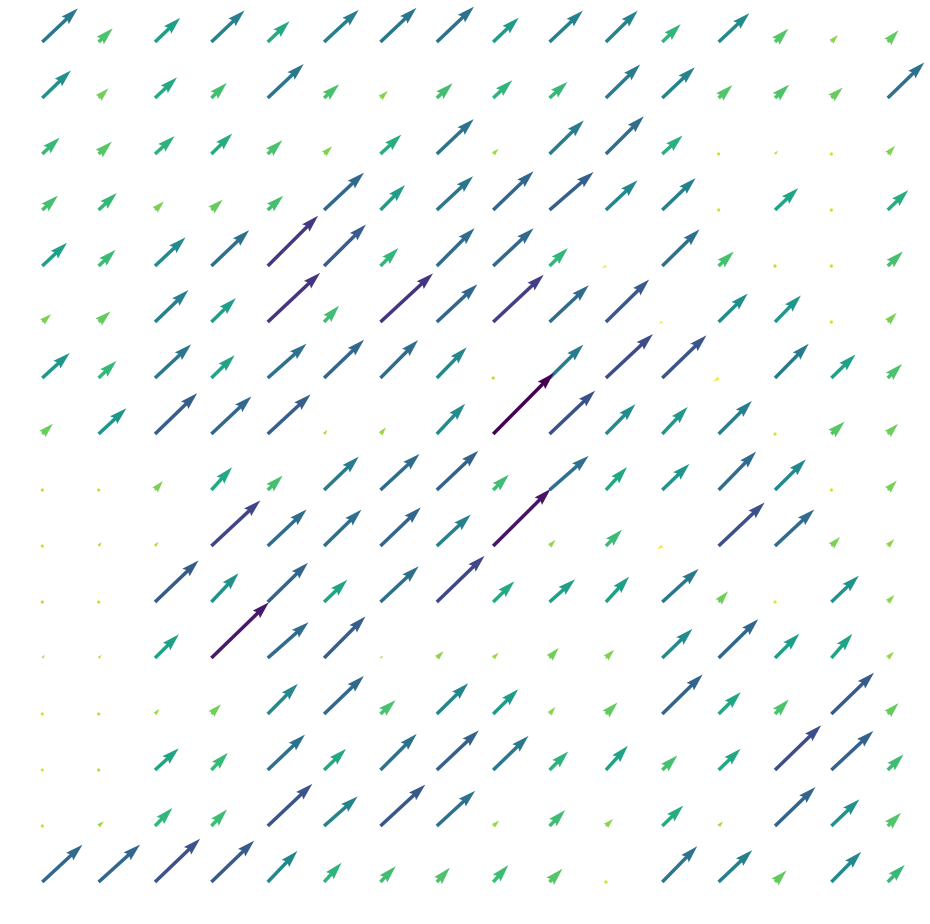}
     & \includegraphics[width=0.24\linewidth,cframe=black]{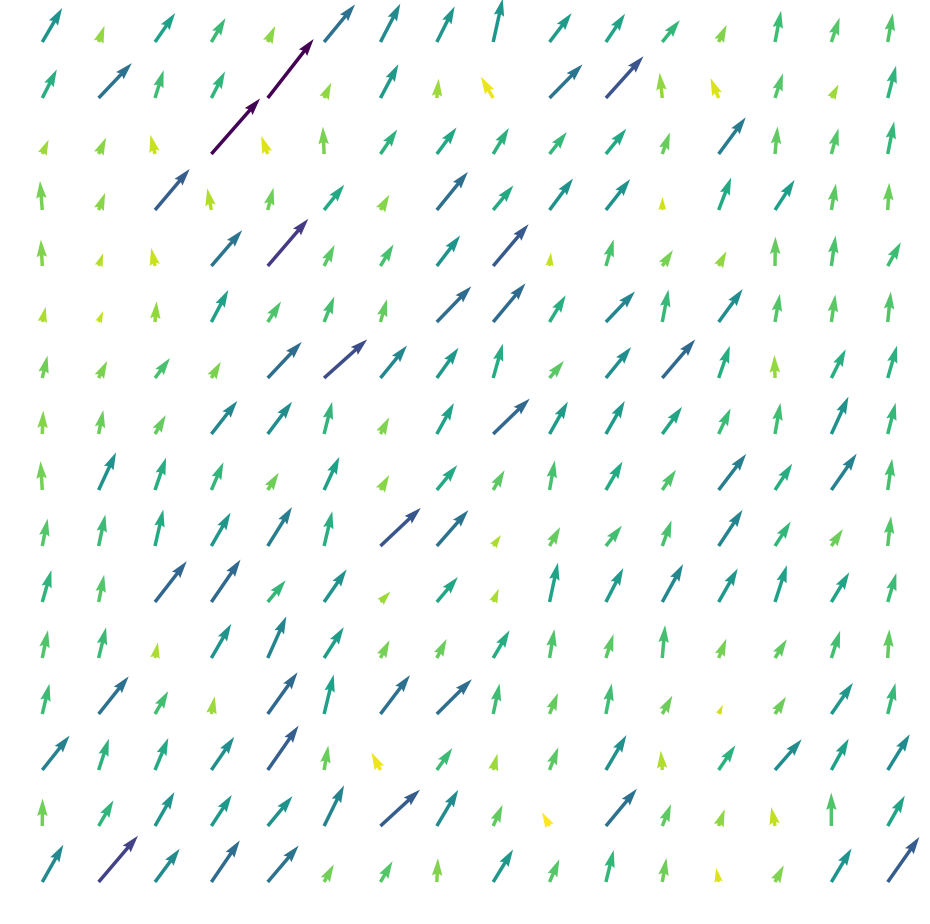}
     & \includegraphics[width=0.24\linewidth,cframe=black]{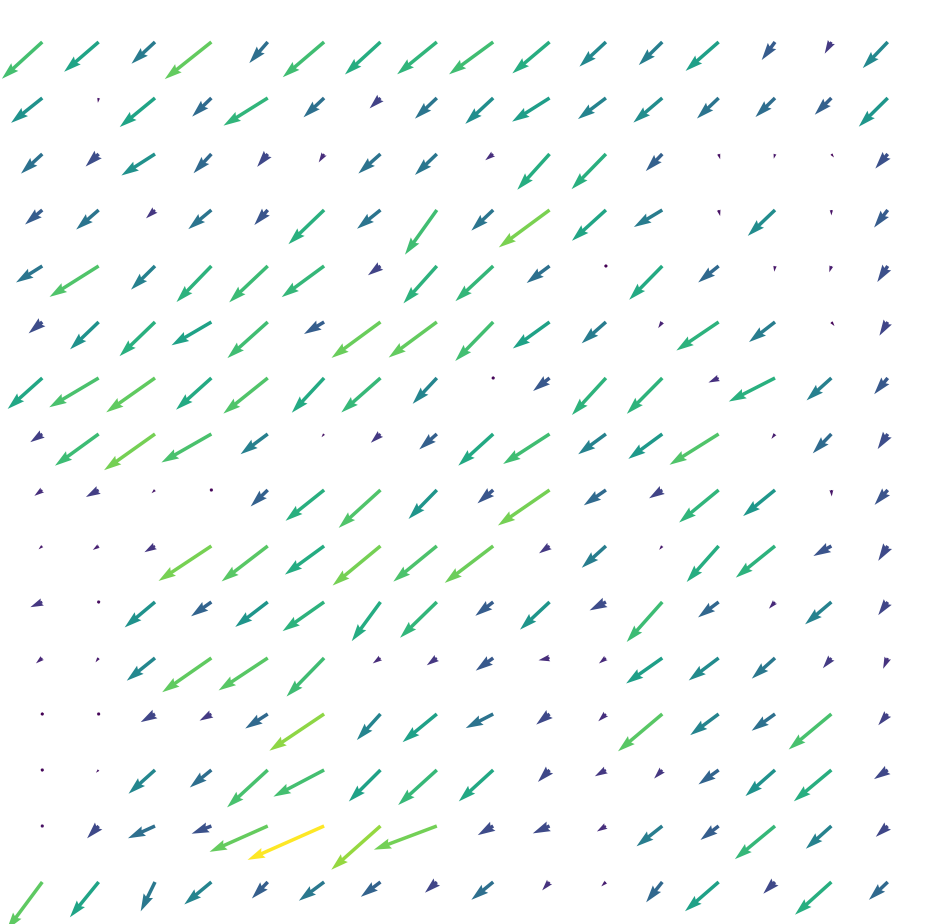}
     \\
     Input & offset $r=4$ & offset $r=8$ & offset $r=16$
\end{tabular}
    \caption{Visualization of varied offsets. The offsets are generated according to the Input image. The sampling step is 16.}
    \label{fig:offsets}
\end{figure}

\noindent\textbf{Effects of offsets $\Delta p$}. In \cref{tab:offset}, we validate the performance changes to explore the effectiveness of offsets by adjusting the offset scalar $r$ from 0 to 16. It can be observed
that the model with $r$ = 8 performs best. In contrast, when $r$ is set to 4 or 16, the model performance negligibly improves but even drops. To comprehensively understand, we depict the offset vector in \cref{fig:offsets}. Offsets with $r$ = 8 are more reasonable since the most effective shifts are located at the edge areas and provide fine-grained partition.

\begin{table}
  \centering
  \fontsize{8.5pt}{9.5pt}\selectfont
    \tabcolsep=4.5pt
  \caption{Ablation study on offsets.}
  \begin{tabular}{c|cccccc}
    \whline
    Offsets Scalar $r$ & \multicolumn{2}{c}{B100} & \multicolumn{2}{c}{Urban100} & \multicolumn{2}{c}{Manga109} \\
    \whline
    0 (\emph{No offsets}) & 27.69 & \textcolor{gray}{$\Delta$} & 26.60 & \textcolor{gray}{$\Delta$}& 31.10& \textcolor{gray}{$\Delta$}\\
    1 & 27.71 & \textcolor{gray}{+0.02} & 26.60 & \textcolor{gray}{+0.00} & 31.14 & \textcolor{gray}{+0.04}\\
    4 & 27.70 & \textcolor{gray}{+0.01} & 26.55 & \textcolor{gray}{-0.05} & 31.14 & \textcolor{gray}{+0.04}\\
    8 & \red{27.72} & \textcolor{gray}{\red{+0.03}} & \red{26.63} & \textcolor{gray}{\red{+0.03}} & \red{31.18} & \textcolor{gray}{\red{+0.08}}\\
    16 & 26.68 & \textcolor{gray}{-0.01} & 26.51 & \textcolor{gray}{-0.09} & 31.06 & \textcolor{gray}{-0.04} \\
    \whline
  \end{tabular}
  \label{tab:offset}
\end{table}

\noindent\textbf{Effects of input conditions}. We ablate the input conditions (local, global, and window) in \cref{tab:condition}. Since our method is content-aware, we maintain the local content for all models. The absence of the global/window condition will cause 0.08dB drops on Urban100. Disabling both of them further decreases the PSNR to 26.46dB, which is 0.17dB lower than the initial model. The results indicate the combinations of multiple conditions can improve the predictor accuracy.

\begin{table*}[!t]\scriptsize
\center
\begin{center}
\caption{Quantitative comparison (PSNR) on 2K-8K testsets. The {\sethlcolor{tablegreen}\hl{accelerating framework}}, \sethlcolor{tablered}\hl{lightweight models}, and \sethlcolor{tableblue}\hl{proposed network} are reported for general comparison. For \algname{}, the ``+ ClassSR'' uses {\algname{}}-S/M/B-\emph{Original} as three branches with different complexities. For mixed strategies of Content-Aware and ClassSR, ``{\color{red}{$\uplus$ {ClassSR}}}'' uses \algname{}-B ($\gamma=0.25/0.3/0.5$) as three branches for ClassSR. ``{\color{red}{$\uplus$ \emph{CAMixer}}}'' represent using \algname{}-S/M/B ($\gamma=0.5$) for ClassSR. For ClassSR-like method, the \{tile, overlap\} is \{32, 2\}, while for lightweight models,  the \{tile, overlap\} is \{64, 4\}, to maintain the max FLOPs under 5G.}

\label{tab:large_input_sr_results}
\small
\footnotesize
\fontsize{8.5pt}{9.5pt}\selectfont
\tabcolsep=2pt
\begin{tabular}{llc|cc|cc|cc|cc}
\whline
 {Method} &   & {\#Params} & F2K & {\#FLOPs}  & Test2K & {\#FLOPs} & Test4K & {\#FLOPs}  & Test8K & {\#FLOPs}
\\
\whline
& \emph{Original}~\cite{SRGAN} & 1.5M
& 29.01
& 5.20G\,\gray{(100\%)}
& 26.19
& 5.20G\,\gray{(100\%)}
& 27.65
& 5.20G\,\gray{(100\%)}
& 33.50
& 5.20G\,\gray{(100\%)}
\\
& \,\,\,+ {\sethlcolor{tablegreen}\hl{ClassSR}}~\cite{ClassSR} & 3.1M
& 29.02
& 3.43G\,\gray{(66\%)} 
& 26.20
& 3.62G\,\gray{(70\%)}
& 27.66
& 3.30G\,\gray{(63\%)}
& 33.50
& 2.70G\,\gray{(52\%)}
\\
& \,\,\,+ {\sethlcolor{tablegreen}\hl{ARM-L}}~\cite{ARM}\quad\quad & 1.5M
& 29.03
& 4.23G\,\gray{(81\%)}
& 26.21
& 4.00G\,\gray{(77\%)}
& 27.66
& 3.41G\,\gray{(66\%)}
& 33.52
& 3.24G\,\gray{(62\%)}
\\
& \,\,\,+ {\sethlcolor{tablegreen}\hl{ARM-M}}~\cite{ARM}\quad\quad & 1.5M
& 29.01
& 3.59G\,\gray{(69\%)}
& 26.20
& 3.48G\,\gray{(67\%)}
& 27.65
& 3.24G\,\gray{(62\%)}
& 33.50
& 2.47G\,\gray{(48\%)}
\\
\multirow{-4}{*}{{SRResNet}}
& \,\,\,+ {\sethlcolor{tablegreen}\hl{ARM-S}}~\cite{ARM}\quad\quad & 1.5M
& 28.97
& 2.74G\,\gray{(53\%)}
& 26.18
& 2.87G\,\gray{(55\%)}
& 27.63
& 2.77G\,\gray{(53\%)}
& 33.46
& 1.83G\,\gray{(35\%)}
\\
& \,\,\,+ {\sethlcolor{tableblue}\hl{\emph{CAMixer}}}\quad\quad & 925K
& 29.15 
& 3.24G 
& 26.31
& 3.24G 
& -
& -
& -
& -
\\
\hline
 & \emph{Original}~\cite{RCAN} & 15.6M
& -
& -
& 26.39
& 32.60G\,\gray{(100\%)}
& 27.89
& 32.60G\,\gray{(100\%)}
& 33.76
& 32.60G\,\gray{(100\%)}
\\
\multirow{-2}{*}{RCAN} 
& \,\,\,+ {\sethlcolor{tablegreen}\hl{ClassSR}}~\cite{ClassSR} & 30.1M
& -
& -
& 26.39
& 21.22G\,\gray{(65\%)}
& 27.88
& 19.49G\,\gray{(60\%)}
& 33.73
& 16.36G\,\gray{(50\%)}
\\
\whline
IMDN & {\sethlcolor{tablered}\hl{\emph{Original}}}~\cite{IMDN} 
& 715K 
& 29.03
& 1.46G
& 26.19
& 1.46G
& 27.65
& 1.46G
& 33.57
& 1.46G
\\
\hline
SwinIR-light & {\sethlcolor{tablered}\hl{\emph{Original}}}~\cite{SwinIR} 
& 930K 
& 29.24
& 2.10G
& 26.33
& 2.10G
& 27.79
& 2.10G
& 33.67
& 2.10G
\\
\whline
& {\sethlcolor{tableblue}\hl{\emph{Original}}}    & {351K} 
& 29.12
& 894M\,\gray{(100\%)}
& 26.26
& 894M\,\gray{(100\%)}
& 27.73
& 894M\,\gray{(100\%)}
& 33.66
& 894M\,\gray{(100\%)}
\\
\multirow{-2}{*}{{\textbf{\algname{}}-S}} 
& \,\,\,+ {\sethlcolor{tableblue}\hl{\emph{CAMixer}}}   &  {351K}
& 29.08
& 652M\,\gray{(73\%)}
& 26.24
& 652M\,\gray{(73\%)}
& 27.70
& 652M\,\gray{(73\%)}
& 33.63
& 652M\,\gray{(73\%)}
\\
\hline
&  {\sethlcolor{tableblue}\hl{\emph{Original}}}  & {535K}
& 29.20
& 1.37G\,\gray{(100\%)}
& 26.32
& 1.37G\,\gray{(100\%)}
& 27.80
& 1.37G\,\gray{(100\%)}
& 33.72
& 1.37G\,\gray{(100\%)}
\\
\multirow{-2}{*}{{\textbf{\algname{}}-M}}
&  \,\,\,+ {\sethlcolor{tableblue}\hl{\emph{CAMixer}}} & {535K}
& 29.18
& 1.03G\,\gray{(75\%)}
& 26.30
& 1.03G\,\gray{(75\%)}
& 27.79
& 1.03G\,\gray{(75\%)}
& 33.71
& 1.03G\,\gray{(75\%)}
\\
\hline
& {\sethlcolor{tableblue}\hl{\emph{Original}}} & {765K} 
& 29.31
& 1.96G\,\gray{(100\%)}
& 26.39
& 1.96G\,\gray{(100\%)}
& 27.89
& 1.96G\,\gray{(100\%)}
& 33.81
& 1.96G\,\gray{(100\%)}
\\
&  \,\,\,+ {\sethlcolor{tableblue}\hl{\emph{CAMixer}}}  & {765K}
& 29.30
& 1.49G\,\gray{(76\%)}
& 26.38
& 1.49G\,\gray{(76\%)}
& 27.87
& 1.49G\,\gray{(76\%)}
& 33.81
& 1.49G\,\gray{(76\%)}
\\
{\textbf{\algname{}}-B}
& \,\,\,\,\,\, {\color{red}{$\uplus$ {\sethlcolor{tableblue}\hl{ClassSR}}}}~\cite{ClassSR}
&  820K
& 29.19
& 1.35G\,\gray{(69\%)}
& 26.32
& 1.37G\,\gray{(70\%)}
& -
& -
& -
& -
\\
& \,\,\,+ {\sethlcolor{tableblue}\hl{ClassSR}}~\cite{ClassSR} &  1711K
& 29.18
& 1.44G\,\gray{(73\%)}
& 26.28
& 1.52G\,\gray{(77\%)}
& 27.76
& 1.53G\,\gray{(78\%)}
& 33.65
& 1.48G\,\gray{(75\%)}
\\
& \,\,\,\,\,\, \color{red}{$\uplus$ {\sethlcolor{tableblue}\hl{\emph{CAMixer}}}}
&  1711K
& 29.17
& 1.04G\,\gray{(53\%)}
& 26.26
& 1.11G\,\gray{(57\%)}
& -
& -
& -
& -
\\
\whline
\end{tabular}
\end{center}
\end{table*}

\begin{table*}[!t]
\center
\small
\footnotesize
\fontsize{8.5pt}{9.5pt}\selectfont
\tabcolsep=4.2pt
\begin{center}
\caption{Quantitative comparison (average PSNR/SSIM, Parameters, and Mult-Adds) with state-of-the-art approaches for efficient image SR.
The best results and the second best results are in \red{bold} and \blue{underline}, respectively. Mult-Adds (MAdds) are measured under the setting of upscaling the image to 1280$\times$720. More results are available in the \emph{supplementary material}.} 
\label{tab:lightweight_sr_results}

\begin{tabular}{l|c|cc|c|c|c|c|c|c|c|c|c|c}
\whline
\multirow{2}{*}{Method} & \multirow{2}{*}{Scale} & \multirow{2}{*}{\#Params} & \multirow{2}{*}{\#MAdds}&  \multicolumn{2}{c|}{Set5~\cite{Set5}} &  \multicolumn{2}{c|}{Set14~\cite{Set14}} &  \multicolumn{2}{c|}{BSD100~\cite{B100}} &  \multicolumn{2}{c|}{Urban100~\cite{Urban100}} &  \multicolumn{2}{c}{Manga109~\cite{manga109}}
\\
\cline{5-14}
& &   &  & PSNR & SSIM & PSNR & SSIM & PSNR & SSIM & PSNR & SSIM  & PSNR & SSIM 
\\
\whline
IMDN~\cite{IMDN} & $\times$4  & 715K & 40.9G
& 32.21
& 0.8948
& 28.58
& 0.7811
& 27.56
& 0.7353 
& 26.04
& 0.7838
& 30.45
& 0.9075
\\
LatticeNet~\cite{LatticeNet}  & $\times$4  & 777K & 43.6G
& {32.18}
& {0.8943}
& {28.61}
& {0.7812}
& {27.57}
& {0.7355}
& {26.14}
& {0.7844}
& -
& -
\\
FDIWN~\cite{FDIWN} & $\times$4  & 664K & 28.4G
& {32.23}
& {0.8955}
& {28.66}
& {0.7829}
& {27.62}
& {0.7380}
& {26.28}
& {0.7919}
& -
& -
\\
SwinIR-light~\cite{SwinIR} & $\times$4  & 930K & {61.7G} %
&  {32.44}
&  {0.8976}
& {28.77}
&  {0.7858}
&  {27.69}
&  {0.7406}
& {26.47}
& {0.7980}
& 30.92 
& 0.9151
\\
ELAN-light~\cite{ELAN} & $\times$4  & 601K & 43.2G
& {32.43}
& {0.8975}
&  {28.78}
&  {0.7858}
&  {27.69}
&  {0.7406}
&  {26.54}
&  {0.7982}
& 30.92
& 0.9150
\\
NGswin~\cite{NGSwin} & $\times$4  & 1019K & {36.4G}
&  {32.33}
&  {0.8963}
&  {28.78}
&  {0.7859}
&  {27.66}
&  {0.7396}
&  {26.45}
&  {0.7963}
& 30.80 
& 0.9128
\\
SwinIR-NG~\cite{NGSwin} & $\times$4  & 1201K & {63.0G}
&  {32.44}
&  {0.8980}
&  \red{28.83}
&  {0.7870}
&  {27.71}
&  {0.7411}
&  {26.54}
&  {0.7998}
& 31.09
& 0.9161
\\
DiVANet~\cite{DiVANet} & $\times$4  & 939K & {57.0G}
& {32.41}
& {0.8973}
& {28.70}
& {0.7844}
& {27.65}
& {0.7391}
& {26.42}
& {0.7958}
& 30.73 
& 0.9119
\\
\rowcolor{tableblue} 
\textbf{\algname{}}\quad\quad\quad\quad  & $\times$4  & {765K} & {53.8G} %
& {\red{32.51}}
& {\red{0.8988}}
& {\blue{28.82}}
& {\red{0.7870}}
& {\red{27.72}}
& {\red{0.7416}}
& {\red{26.63}}
& {\red{0.8012}}
& {\red{31.18}}
& {\red{0.9166}}
\\
\whline 
\end{tabular}
\end{center}
\end{table*}

\subsection{Large-Image SR}
Following ClassSR~\cite{ClassSR} and ARM~\cite{ARM}, we validate the efficiency of \algname{} with 2K-8K large-image SR task. Unlike previous work omitting lightweight models, we also add IMDN~\cite{IMDN} and SwinIR-light~\cite{SwinIR} for reference. All models are running under the same FLOPs restriction.

\begin{table*}[!t]\scriptsize
\center
\begin{center}
\caption{Quantitative comparison on ODI-SR, SUN 360, under Fisheye downsampling. ``$\dagger$'' represents using augmentation dataset.}

\label{tab:odisr_results}
\small
\footnotesize
\fontsize{8.5pt}{9.5pt}\selectfont
\tabcolsep=6pt
\begin{tabular}{l|c|c|cccc|cccc}
\whline
\multirow{2}{*}{Method} & \multirow{2}{*}{Scale} & \multirow{2}{*}{\#Params} & \multicolumn{4}{c|}{ODI-SR~\cite{LAU-Net}} & \multicolumn{4}{c}{SUN 360 Panorama~\cite{sun360}}
\\
\cline{4-11}
& & & PSNR & SSIM & WS-PSNR & WS-SSIM & PSNR & SSIM & WS-PSNR & WS-SSIM \\
\whline
Bicubic & $\times$2 & -
& 28.21
& 0.8215
& 27.61
& 0.8156
& 28.14
& 0.8118
& 28.01
& 0.8321
\\
RCAN~\cite{RCAN} & $\times$2 & 15.4M
& 30.08
& 0.8723
& 29.49
& 0.8714 
& 30.56
& 0.8712
& 31.18
& 0.8969
\\
SRResNet~\cite{SRGAN} & $\times$2 & -
& 30.16
& 0.8717
& 29.59
& 0.8697 
& 30.64
& 0.8714
& 31.20
& 0.8953
\\
EDSR~\cite{EDSR} & $\times$2 & 40.7M
& 30.32 
& 0.8770
& 29.68
& 0.8727 
& 30.89
& 0.8784
& 31.42
& 0.8995
\\
OSRT-light$^\dagger$~\cite{OSRT} \quad\quad & $\times$2 & 1.26M
& 30.42
& 0.8775
& 29.79
& 0.8735 
& 31.00
& 0.8792
& 31.55
& 0.9004
\\
\rowcolor{tableblue}
\textbf{\algname{}} & $\times$2 & 1.30M
& \red{30.46} 
& \red{0.8789}
& \red{29.83}
& \red{0.8747}
& \red{31.04}
& \red{0.8810}
& \red{31.60}
& \red{0.9020}
\\
\whline
Bicubic & $\times$4 & -
& 25.59
& 0.7118 
& 24.95 
& 0.6923 
& 25.29 
& 0.6993 
& 24.90 
& 0.7083
\\
RCAN~\cite{RCAN} & $\times$4 & 15.6M
& 26.85
& 0.7621
& 26.15
& 0.7485 
& 27.10
& 0.7660
& 26.99
& 0.7856
\\
SRResNet~\cite{SRGAN}  & $\times$4 & -
& 26.91
& 0.7597
& 26.24
& 0.7457
& 27.10 
& 0.7618  
& 26.99 
& 0.7812
\\
EDSR~\cite{EDSR} & $\times$4 & 43.1M
& 26.97 
& 0.7589
& 26.30 
& 0.7458
& 27.19
& 0.7633
& 27.10
& 0.7827
\\
OSRT-light$^\dagger$~\cite{OSRT} & $\times$4 & 1.28M
& 27.17
& 0.7667
& 26.49
& 0.7526
& 27.48
& 0.7718
& \red{27.41}
& 0.7911
\\
\rowcolor{tableblue}
\textbf{\algname{}} & $\times$4 & 1.32M
& \red{27.19}
& \red{0.7691}
& \red{26.49}
& \red{0.7538}
& \red{27.48}
& \red{0.7736}
& {27.36}
& \red{0.7916}
\\
\whline
\end{tabular}
\end{center}
\end{table*}

\begin{figure*}
    \centering
    \scriptsize
    \tabcolsep=2pt
    \begin{tabular}{cccccc}
         \multirow{-7}{*}{\includegraphics[width=0.38\linewidth,height=0.255\linewidth]{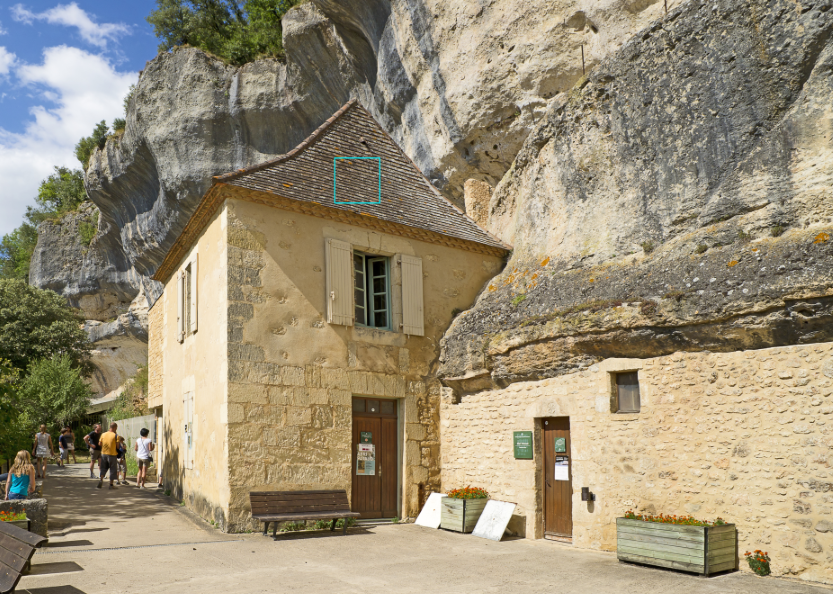}}
         & \includegraphics[width=0.11\linewidth,height=0.11\linewidth]{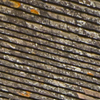} 
         & \includegraphics[width=0.11\linewidth,height=0.11\linewidth]{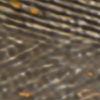} 
         & \includegraphics[width=0.11\linewidth,height=0.11\linewidth]{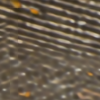}
         & \includegraphics[width=0.11\linewidth,height=0.11\linewidth]{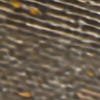} 
         & \includegraphics[width=0.11\linewidth,height=0.11\linewidth]{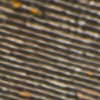} 
         \\
         & \multirow{2}{*}{\small HR} & SRResNet-\emph{O}~\cite{SRGAN} & RCAN-\emph{O}~\cite{RCAN} & IMDN-\emph{O}~\cite{IMDN} & \textbf{\algname{}}-\emph{O} \\
         & &  16.69/5.20G & 17.26/32.60G & 17.26/1.46G & 18.99/1.96G \\
         & \includegraphics[width=0.11\linewidth,height=0.11\linewidth]{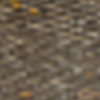} 
         & \includegraphics[width=0.11\linewidth,height=0.11\linewidth]{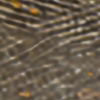} 
         & \includegraphics[width=0.11\linewidth,height=0.11\linewidth]{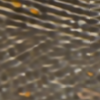}
         & \includegraphics[width=0.11\linewidth,height=0.11\linewidth]{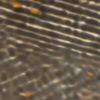} 
         & \includegraphics[width=0.11\linewidth,height=0.11\linewidth]{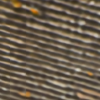} 
         \\
         \multirow{2}{*}{\small \emph{1310} from Test4K} & \multirow{2}{*}{\small LR} & SRResNet-\emph{ClassSR} & RCAN-\emph{ClassSR} & SwinIR-L-\emph{O}~\cite{SwinIR} & \textbf{\algname{}} \\
         & &  15.88/3.43G & 16.56/21.22G & 16.69/2.10G & 18.83/1.49G 
         \\
    \end{tabular}
    \caption{Visual comparison of \algname{} with other methods for $\times$4 task on Test4K dataset.}
    \label{fig:Test4K}
\end{figure*}

\noindent\textbf{Quatitative results}.
In \cref{tab:large_input_sr_results}, we implement \algname{}-\textbf{S}mall/\textbf{M}edium/\textbf{B}ase with 36/48/60 channels. Generally, \algname{} remarkably surpasses other methods. Against classic SRResNet and RCAN accelerated by ClassSR~\cite{ClassSR} or ARM~\cite{ARM}, the \algname{}-\emph{Original} delivers similar or better restoration quality but is 3.2$\times$-11$\times$ lighter. Moreover, our \emph{CAMixer} models further save about 25\% calculations. Overall, \algname{} (765K/747M) can compete with RCAN (15.6M/32.6G) for 2K-8K image restoration. In comparison with lightweight models SwinIR-light, our \algname{} renders 0.14dB PSNR improvement with fewer FLOPs or 51\% fewer FLOPs with higher PSNR.

To fairly compare the accelerating framework, \ie, ClassSR with the proposed \emph{Content-Aware} Mixing, we apply ClassSR to \algname{} with three classes. The ClassSR helps to reduce similar computations as \emph{Content-Aware} but falls behind 0.12dB on F2K. This PSNR gap may result from to limited receptive field of small cropping size, which we will explore in further research. We also illustrate two potential combination ways for ClassSR and CAMixer. {\color{red}{``$\uplus$\,ClassSR"}} indicates applying ClassSR on \algname{}-B, \ie, using the same backbone but adjusting $\gamma=0.25/0.3/0.5$ as three branches for ClassSR. {\color{red}{``$\uplus$\,CAMixer"}} employs CAMixer with $\gamma=0.5$ for \algname{}-S/M/B  as three branches of ClassSR.
As expected, the former approach maintains better restoration quality while the latter reduces more calculations. Conclusively, \emph{Content-Aware} mixer is a better choice for large-image tasks and can work with other strategies, without any difficulty.

\noindent\textbf{Qualitaive results}. In \cref{fig:Test4K}, we present the visual results of the \algname{} against other methods. \algname{} recovers significantly clearer lattice content than other approaches. Moreover, the \algname{} with {Content-Aware} strategy induces less performance drop than ClassSR.

\subsection{Lightweight SR}
To evaluate the generality of CAMixer, we compare the proposed \algname{} with numerous SOTA lightweight models, including IMDN~\cite{IMDN}, LatticeNet~\cite{LatticeNet}, FDIWN~\cite{FDIWN}, SwinIR-light~\cite{SwinIR}, ELAN-light~\cite{ELAN}, NGswin~\cite{NGSwin}, and DiVANet~\cite{DiVANet}. \cref{tab:lightweight_sr_results} shows the quantitative comparison. Inclusively, our \algname{} obtains superior restoration quality on all five benchmark datasets with moderate parameters and fewer computations. In particular, compared to the newest SwinIR-NG~\cite{NGSwin} using self-attention for all tokens, the \algname{} can attain better performance while saving 9.2G computations. The results strongly indicate that our CAMixer can work not only for large input images with plenty of plain areas but also for classic SR tasks.

\subsection{Omni-Directional-Image SR}
To understand the effectiveness of CAMixer under practical utilization with large-resolution inputs, we test our \algname{} on the Omni-Directional-Image (ODI) SR task, which has 2K output and complex distortion. Specifically, we add the distortion
map as an extra condition for predictors.  In \cref{tab:odisr_results}, we exhibit the qualitative comparison of our \algname{} with other methods. Our \algname{} obtains better restoration quality for almost all validation metrics. In detail, compared with 30$\times$ larger EDSR~\cite{EDSR}, \algname{} obtains 0.26dB improvement on SUN 360 benchmark. Moreover, \algname{} advances OSRT-light~\cite{OSRT} by a maximum of 0.05dB/0.0028 (WS-PSNR/WS-SSIM) improvements without using additional training sets. 

%% file: sec/3_finalcopy.tex
\section{Conclusion}

In this paper, we propose a content-aware mixer (CAMixer) that integrates the model accelerating scheme and token mixer design by routing neural operators (self-attention and convolution) of varied complexities according to the difficulty of content recovery. Particularly, the simple tokens are captured by convolution while the complex tokens are additionally processed by deformable self-attention. To improve the accuracy of routing, we introduce an effective predictor, which uses rich input conditions to generate multiple useful guiding information. Based on CAMixer, we build \algname{}, which reaches remarkable performance-calculation trade-offs on three SR tasks. 